\documentclass[prd,aps,superscriptaddress,11pt,nobibnotes,notitlepage,longbibliography,nofootinbib,preprintnumbers]{revtex4-2}
 
\usepackage[normalem]{ulem}
\usepackage{sidecap}
\usepackage[latin1]{inputenc}
\usepackage{graphicx}
\usepackage{float}
\usepackage{latexsym}
\usepackage{graphicx}
\usepackage{amssymb}
\usepackage{amsmath,mathtools}
\usepackage{amsfonts}
\usepackage{tikz} 
\usepackage{ragged2e}
\usepackage{stackengine}
\usepackage{braket}
\usepackage{scalerel}
\usepackage{MnSymbol}

\usepackage[normalem]{ulem}
\usepackage{accents}
\usepackage{tikz} 
\usepackage[margin=2cm]{geometry}

\usepackage[
    colorlinks=true,
    citecolor=red,
    linkcolor=blue,
    urlcolor=black,
    hyperfootnotes=false
]{hyperref}

\usepackage{dcolumn}
\usepackage{textcomp}
\usepackage[dvipsnames]{xcolor}
\usepackage{xfrac}
\usepackage{slashed}
\usepackage{multirow}
\renewcommand{\d}{{\rm d}}
\newcommand{\intd}{\int {\rm d}}

\def\be{\begin{equation}}
\def\ee{\end{equation}}
\def\figs/B{B}
\def\bea{\begin{eqnarray}}
\def\eea{\end{eqnarray}}
\def\bg{\begin{eqnarray}}
\def\nd{\end{eqnarray}}
\def\sin{{\rm sin}}
\def\cos{{\rm cos}}

\def\log{{\rm log}}
\def\ln{{\rm log}}

\def\beq{\begin{equation}}
\def\eeq{\end{equation}}

\definecolor{oiOrange}{RGB}{230,159,0}
\definecolor{oiBlue}{RGB}{0,114,178}
\definecolor{oiGreen}{RGB}{0,158,115}

\usepackage{graphicx}
\usepackage{dcolumn}
\usepackage{bm}
\usepackage{hyperref}
\usetikzlibrary{decorations.pathmorphing}
\tikzset{snake it/.style={decorate, decoration=snake}}

\begin{document}

\preprint{IFT-UAM/CSIC-26-44}

\title{On quantum tunnelling in the presence of Noether charges
}

\author{Giulio Barni}
 \email{giulio.barni@ift.csic.es}
\affiliation{Instituto de F\'isica Te\'orica IFT-UAM/CSIC, Cantoblanco, E-28049, Madrid, Spain}%
\author{Thomas Steingasser}
 \email{thomas.steingasser@uam.es}
 \affiliation{Instituto de F\'isica Te\'orica IFT-UAM/CSIC, Cantoblanco, E-28049, Madrid, Spain}%
\affiliation{%
Departamento de Fisica Teorica, Universidad Autonoma de Madrid}

\date{\today}

\begin{abstract}
We provide a complete first-principles based discussion of quantum tunnelling out of initial states carrying a conserved Noether charge. Our main result is a simple, unambiguous Euclidean-time prescription for the calculation of tunnelling rates out of such states. By relying on a combination of the direct approach and the steadyon framework for the evaluation of real-time path integrals, our derivation offers full transparency of its underlying assumptions, and is independent of any ad-hoc generalisations. This strategy also offers a simple explanation for the emergence of complex saddle points for such systems, justifying techniques postulated by earlier works. Our analysis furthermore offers the first results for initial states with both a conserved Noether charge and a non-trivial energy.\\
We first illustrate the main conceptual points of our analysis for the simple example of a point particle in two spatial dimensions carrying a conserved angular momentum. Then, we generalise our results to the case of multiple dimensions and an arbitrary conserved Noether charge, providing an easy-to-implement prescription for the calculation of the tunnelling rate. We furthermore apply our results to the example of a complex scalar field subject to a global $U(1)$-symmetry with associated charge. These results provide a reliable foundation for the calculation of tunnelling rates in applications in finite-density and charge-asymmetric systems.
\end{abstract}

\maketitle

\vspace{-0.5cm}

\tableofcontents

\section{Introduction}

Quantum tunnelling is one of the most characteristic non-perturbative phenomena in all quantum theories. In quantum mechanics, tunnelling describes the exponentially suppressed propagation of a particle through a classically forbidden region, providing the basic mechanism behind phenomena such as $\alpha$ decay and the splitting of quasi-degenerate levels in double-well potentials \cite{Gamow:1928zz,Gurney:1928,Coleman:1977py,PhysRevD.16.1762}. In field theory, it underlies metastable vacuum decay and false-vacuum transitions, with applications ranging from particle physics to cosmology \cite{Coleman:1977py,PhysRevD.16.1762,Coleman:1980aw,LINDE198137}. These processes have been extensively studied in the literature, leading to a robust theoretical understanding of tunnelling out of the simplest false vacuum states in terms of a Euclidean-time picture~\cite{Andreassen:2016cvx}.

However, in a broad range of applications, one is interested in decay processes occurring at finite energies or fixed values of conserved quantities, such as angular momentum in quantum mechanics or Noether charges associated with a continuous symmetry in quantum field theory \cite{Lee:1994np,Kao:1995gf,Levkov:2017paj,Barni:2026toappear}.
As noted in several of these earlier works, a generalisation of the familiar Euclidean-time picture that properly incorporates a non-zero Noether charge appears to necessitate \textit{complex saddle points}, in contrast with the strictly real instantons encountered in the usual treatment of false-vacuum tunnelling~\cite{Lee:1988zz,Coleman:1990as,Lee:1994np,Levkov:2017paj,Barni:2026toappear}.

This behaviour can be traced back to the properties of the Noether charges themselves. In contrast to the energy, conserved charges of this kind generically become complex after Wick rotation, and therefore cannot be implemented in a Euclidean description built solely from strictly real degrees of freedom~\cite{Lee:1988zz,Giddings:1988cx,Giddings:1988wv,Coleman:1990as,Lee:1994np,Alonso:2017avz,Loges:2022zxn}. Consider, for instance, a conserved Noether charge $C$ associated to a symmetry along some generalised coordinate $\theta$. Following Noether's theorem, $C$ can be obtained from the Lagrange function $L$ through the simple relation
\begin{align}
    C= \frac{\partial L}{\partial \dot{\theta}}\propto \dot{\theta}\, .
\end{align}
Thus, $C$ transforms under a Wick rotation $t\to -i \tau$ as $C (\dot{\theta})\to i C (\d \theta/\d \tau)$.

Crucially, this generally requires a genuine complexification of the relevant degrees of freedom, rather than a mere small complex deformation. In particular, these works have found that the presence of a conserved Noether charge forces the instanton component along the cyclic variable to be strictly imaginary~\cite{Lee:1994np,Levkov:2017paj,Barni:2026toappear}. Beyond the formal questions this raises, such a complexification also introduces the possibility of additional saddle points. If, a priori, \textit{any} complex saddle point were allowed to contribute to the relevant path integrals, this could lead to a large proliferation of new solutions, potentially with a significant impact on the tunnelling rate.

A natural framework to understand the emergence of complex saddle points is offered by the recently developed \textit{steadyon picture}~\cite{Steingasser:2024ikl,Steingasser:2024wkh,Lin:2025bjn,Lin:2025pap} in combination with the direct approach to tunnelling~\cite{Andreassen:2016cff,Andreassen:2016cvx}. In combination, these advances enable us to derive a semi-classical expression for the tunnelling rate from first principles. They allow, in particular, to evaluate the relevant path integrals directly in physical time, from which the familiar Euclidean-time quantities can be obtained through clear, transparent steps. This is made possible through the introduction of a new class of complexified solutions to the (regularised) real-time equations of motion dubbed~\textit{steadyons}. This picture thus offers a natural setup to develop a reliable understanding of the role that complex saddle points seem to play in the description of tunnelling processes in the presence of conserved Noether charges. Thus, the main purpose of this article is the application of this picture to develop a complete first-principles understanding of quantum tunnelling in the presence of such charges, providing a more robust foundation for results that were previously obtained through generalisation or conjecture.

For concreteness, we discuss two particularly illustrative examples. First, we consider a point particle in two spatial dimensions moving in a rotationally invariant potential at fixed angular momentum. This system captures the structural mechanism in its simplest form, with angular momentum playing the role of the conserved Noether charge and the angular variable playing the role of the symmetry direction. Using this simple example, we will demonstrate in great detail how the angular momentum of the system's initial state propagates through to the complex steadyon, ultimately leading to a Euclidean-time description with an imaginary angular variable. Our argument also establishes that this particular saddle point indeed dominates the tunnelling rate, rather than hypothetical other saddle points made feasible through the complexification. We then generalise the emerging Euclidean-time procedure to arbitrary dimensions and present step-by-step instructions for its implementation. Next, we consider the simplest field-theoretic example with a conserved charge, namely a complex scalar theory with global $U(1)$ symmetry, for which we perform a similar analysis. We show that the appearance of complex Euclidean saddles is not a peculiarity of the mechanical example, but a completely general consequence of imposing a non-zero conserved charge in the tunnelling problem. Finally, we discuss the generalisation of these insights to arbitrary tunnelling processes involving conserved Noether charges. This opens the way to applications in dense relativistic matter with one or more conserved charges, including quark-matter nucleation and phase conversion in neutron stars and proto-neutron stars, as well as possible gravitational-wave signatures from first-order QCD phase transitions in core-collapse supernovae and neutron-star mergers \cite{Glendenning:1992vb,Iida:1997ic,Mintz:2009ay,Zha:2020icw,Most:2018eaw,Bauswein:2018bma,Blas:2022xfk,Bea:2024qcdlike,Bai:2025baryondense}.

This work is organised as follows. In Sec.~\ref{sec:FPderivation}, we provide a detailed first-principles based derivation of the tunnelling rate and the probability flux determining it to leading order. For simplicity, we focus on the illustrative example of a point particle in two spatial dimensions, subject to a rotational symmetry giving rise to a conserved angular momentum. This leads us to a compact expression for the tunnelling rate in terms of the theory's propagator. Then, in Sec.~\ref{sec:semi-class}, we evaluate this expression in the semi-classical limit using the steadyon picture. We show, in particular, how the relevant quantities can be obtained directly through a simplified Euclidean-time calculation. We summarise our results in Sec.~\ref{sec:recipe}, providing step-by-step instructions for the leading-order calculation. We furthermore illustrate this procedure, as well as important steps of our derivation, for a concrete example. Next, in Sec.~\ref{sec:steadyon_field_theory}, we discuss the generalisation of our discussion to quantum field theory. For concreteness, we focus on the simplest example of a complex scalar field subject to a global $U(1)$-symmetry. For the readers' convenience, we furthermore include two appendices reviewing the WKB approximation (Appendix~\ref{app:wkb}) and important concepts of the wave functional formulation of quantum field theory (Appendix~\ref{app:config_space_routhian}).

\section{Tunnelling rate and probability flux}
\label{sec:FPderivation}

\begin{figure}
    \centering
    \includegraphics[width=0.49\linewidth]{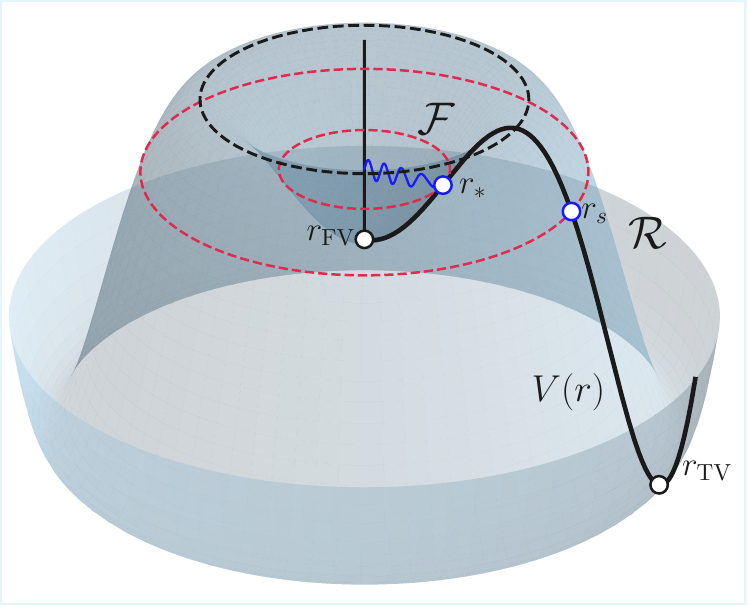}
    \begin{minipage}{0.49\linewidth}
    \centering \vspace{-7cm}
        \includegraphics[width=\linewidth]{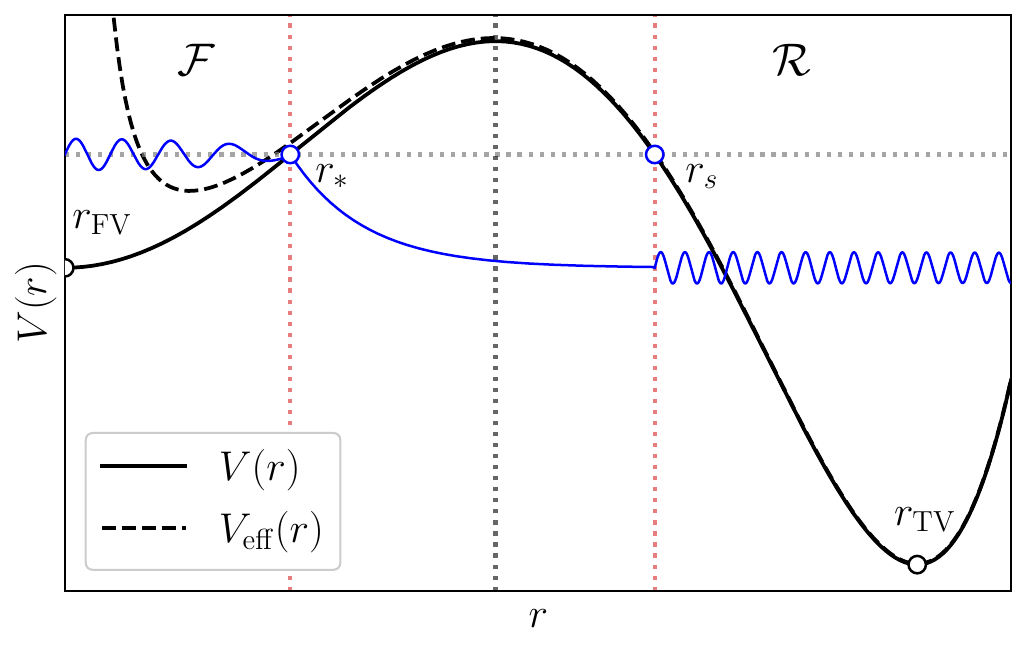}
    \end{minipage}
    \caption{Illustration of the mechanical model. A point particle moves in a rotationally invariant potential \(V(r)\) with a false-vacuum circle at \(r=r_{\rm FV}\) and a lower-lying true-vacuum circle at \(r=r_{\rm TV}\), with \(r_{\rm FV}<r_{\rm TV}\). The left panel shows the corresponding two-dimensional potential landscape, together with the false-vacuum region \(\mathcal{F}\) and the true-vacuum region \(\mathcal{R}\). The right panel shows the radial potential \(V(r)\) (solid black) and the effective potential \(V_{\rm eff}(r)\) at fixed angular momentum (dashed black). The points \(r_\ast\) and \(r_s\) denote the turning points of the reduced radial motion relevant for the fixed-angular-momentum tunnelling problem. The blue oscillatory curves are a schematic representation of the semiclassical wavefunction: oscillatory in the classically allowed regions and exponentially suppressed under the barrier. They are meant only as an illustration of the semiclassical support of the state, and should not be interpreted as the literal time evolution of the particle during tunnelling.}
    \label{fig:sombrero}
\end{figure}

We consider a point particle moving in a rotationally invariant potential in two spatial dimensions, $V=V(r)$. To allow for quantum tunnelling, we assume that this potential has a false vacuum at $r=r_{\rm FV}$ and a lower-lying vacuum at $r=r_{\rm TV}$. For simplicity, we let $r_{\rm FV}<r_{\rm TV}$. We will denote the funnel surrounding the false vacuum circle by $\mathcal{F}$, and its counterpart surrounding the true vacuum by $\mathcal{R}$.\footnote{Note that we will find that the tunnelling rate depends to leading order only on the properties of the barrier between these two regions, and is well-defined even for unbounded potentials.} See Fig.~\ref{fig:sombrero}.

We can now assume the particle to have a definite energy $E$ and angular momentum $J$ and be localised predominantly within the funnel $\mathcal{F}$. This corresponds to a resonance state $|\tilde{E},J\rangle$, where the complex eigenvalue $\tilde{E}$ can be rewritten in terms of a real energy $E$ and tunnelling rate $\Gamma$ as $\tilde{E}=E-i \Gamma/2$. The wave function of such a state is generally of the form $\langle x | \tilde{E},J\rangle\equiv\Psi_{\tilde{E},J}(r,\varphi)=e^{iJ \varphi}\psi_{\tilde{E},J}(r)$. Moving forward we will focus on the semi-classical limit in which $\Gamma \ll E$, allowing us to drop $\Gamma$ from our state labels.

\subsection{First-principles derivation}

To calculate $\Gamma$ directly, we observe that the probability to find the particle within $\mathcal{F}$ changes with time as
\begin{align}
    P_{\mathcal{F}}(t)=\int_\mathcal{F}\d^2 x \ |\langle x| e^{-iHt}| \tilde{E},J \rangle |^2 \simeq \int_\mathcal{F}\d^2 x \ |\langle x|  \tilde{E},J \rangle |^2 e^{-\Gamma t}=e^{-\Gamma t} P_{\mathcal{F}}(0)\, .
\end{align}
Using conservation of probability, this implies that
\begin{align}
    \Gamma = - \frac{\dot{P}_{\mathcal{F}}(t)}{P_{\mathcal{F}}(t)}=\frac{\dot{P}_{\mathcal{R}}(t)}{P_{\mathcal{F}}(t)}\, .
\end{align}
For both $\Omega \in \{\mathcal{F},\mathcal{R} \}$, we can now rewrite these probabilities as
\begin{align}
    P_{\Omega} (t) =& \int_{\Omega} \d^2 x\ \langle x | e^{-i H t} | E,J \rangle \langle E,J | e^{i H t}|x\rangle \nonumber \\ 
    =& \int_{\Omega} \d^2 x \intd x_1\intd x_2\ \Psi_{E,J}(x_1)  \Psi_{E,J}^* (x_2)  \langle x | e^{-i H t} | r_1,\varphi_1 \rangle \langle r_2,\varphi_2 | e^{i H t}|x\rangle \nonumber \\ 
    = & \intd x_1 \intd x_2\ \Psi_{E,J}(x_1) \Psi_{E,J}^* (x_2)\int_{\Omega} \d^2 x  \ D_F(x,t|x_1) D_F^*(x,t|x_2)\, .
\end{align}
Here, as in the remainder of this article, propagators with no explicit time label are to be understood as evaluated at $t=0$.

To evaluate $\dot{P}_{\mathcal{R}}$, we can further generalise the strategy first presented in Ref.~\cite{Andreassen:2016cff} and recently extended to resonance states in Refs.~\cite{Lin:2025bjn,Lin:2025pap}. For the final result, see Eq.~\eqref{eq:PRdotfull}. Following these works, we define the \textit{entrance radius} $r_*$ as well as the \textit{emergence radius} $r_s$ via the conditions
\begin{gather}
    V(r_*)=E=V(r_s), \quad r_* \in \mathcal{F}, \quad r_s \in \mathcal{R}\, .
\end{gather}
Moving forward, we denote the submanifolds with $r=r_*$ and $r=r_s$ by $\Sigma_*$ and $\Sigma_s$, respectively. 

To allow for a straightforward generalisation later on, we first separate the wave functions encoding the initial state from the time-dependent propagators,
\begin{align}
    P_{\mathcal{R}} (t) =&\intd x_1 \intd x_2\ \Psi_{E,J}(x_1) \Psi_{E,J}^* (x_2)\int_{\mathcal{R}} \d^2 x_f  \ D_F(x_f,t|x_1) D_F^*(x_f,t|x_2) \nonumber \\ \label{eq:P_R} 
    \equiv &  \intd x_1 \intd x_2\ \Psi_{E,J}(x_1) \Psi_{E,J}^* (x_2)  p_{\mathcal{R}} (x_1,x_2,t)\, .
\end{align}
To evaluate $p_{\mathcal{R}}$, we furthermore introduce the auxiliary propagators $\bar{D}_F$, $\underaccent{\bar}{D}_F$ and $\bar{\underaccent{\bar}{D}}_F$ as
\begin{align}
    \bar{D}_F (x_s , t_s | x_i) \equiv &  \int_{x (0) = x_i}^{x (t_s) = x_s} {\mathcal D} x \ e^{i S[x]} \delta \left( F_{\Sigma_s}[x] - t_s \right) \quad {\rm for} \quad x_i \in \mathcal{F}\, . \label{eq:Dbar}\\
    \underaccent{\bar}{D}_F ( x_f,t_f | x_* ,t_*) \equiv & \quad \int_{x(t_*) = x_*}^{x(t_f) = x_f} \mathcal{D}x \ e^{iS[x]} \delta (G_{\Sigma_*} [x] -t_*)\quad {\rm for} \quad x_f \in \mathcal{R} \, , \label{eq:Dunderbar} \\ 
    \bar{\underaccent{\bar}{D}}_F ( x_s,t_s | x_* ,t_* )   \equiv & \quad \int_{x(t_*) = x_*}^{x(t_s) = x_s} \mathcal{D}x \ e^{iS[x]} \delta (G_{\Sigma_*} [x] -t_*)  \delta (F_{\Sigma_s} [x] -t_s)  \, . \label{eq:Dbarunderbar}
\end{align}
The functional $F_{\Sigma_s}$ is defined as mapping any path onto the time it \textit{first} crosses $\Sigma_s$, while $G_{\Sigma_*}$ maps any path onto the time it \textit{last} crosses $\Sigma_*$.

These propagators and the original Feynman propagator $D_F$ are related to one another through the decompositions
\begin{align}
	D_F(x_f,t | x_i)= &\int_{0}^{2\pi}\d \varphi_s \int_{0}^t \d t_s D_F( x_f, t |x_s, t_s ) \ \bar{D}_F(x_s,t_s|x_i) \label{eq:Ddecomp1}\\
	D_F(x_s,t | x_i)= &\int_{0}^{2\pi}\d \varphi_* \int_{0}^t \d t_* \underaccent{\bar}{D}_F( x_s, t |x_*, t_* ) \ D_F(x_*,t_*|x_i) \label{eq:Ddecomp2}\\
 \bar{D}_F(x_s,t_s|x_i) = &\int_0^{2 \pi} \d \varphi_* \int_0^{t_s} \d t_* \  \bar{\underaccent{\bar}{D}}_F (x_s,t | x_* ,t_*) D_F(x_*,t_* | x_i)\, ,
\label{eq:Ddecomp3}
\end{align}
where again $x_i \in \mathcal{F}$ and $x_f \in \mathcal{R}$. This form has a clear interpretation: for the particle to move from its initial position $x_i \in \mathcal{F}$ to some $x_f \in \mathcal{R}$, it first needs to cross the submanifold $\Sigma_*$ at some time $t_*$, and then $\Sigma_s$ at some $t_s$.

Closely following Ref.~\cite{Lin:2025bjn}, we can now apply this decomposition to $p_\mathcal{R}$, 
\begin{align}
    p_{\mathcal{R}}(x_1,x_2,t)  = & \int_\mathcal{R}\d^2 x_f\ D_F(x_{f},t|x_1)D_F^*(x_{f},t|x_2) \nonumber \\ 
    =& \int_\mathcal{R}\d^2 x_f \int_0^t \d t_s  \int_{0}^{2\pi}\d \varphi_s  \ D_F ( x_f , t | x_s , t_s) \bar{D}_F (x_s ,t_s | x_1)  \nonumber \\ 
    &\times\int_0^t \d t_s^\prime \int_{0}^{2\pi}\d \varphi_s^\prime  D_F^* (  x_f , t | x_s^\prime , t_s^\prime )  \bar{D}_F^* ( x_s^\prime ,t_s^\prime | x_2 ) \nonumber \\ 
    \simeq & \int_0^t \d t_s  \int_{0}^{2\pi}\d \varphi_s   \bar{D}_F (x_s ,t_s | x_1)   \int_0^t \d t_s^\prime \int_{0}^{2\pi}\d \varphi_s^\prime   \bar{D}_F^* ( x_s^\prime ,t_s^\prime | x_2 ) D_F ( x_s^\prime , t_s^\prime | x_s , t_s)\, ,
\end{align}
where $x_s=\{r_s,\varphi_s\}$ and $x_s^\prime=\{r_s,\varphi_s^\prime\}$. In the last line, we have performed the integral over $x_f$ by assuming back-tunnelling to be subdominant,
\begin{align}
\label{eq:nobacktunnel}
    \int_{\mathcal{R}} \d x_f D_F( x_f, t |x_s, t_s )  D_F^*( x_f, t |x_s^\prime, t_s^\prime ) \simeq D_F( x_s^\prime, t_s^\prime  |x_s, t_s ) \, .
\end{align}
In order to perform the time derivative, we may use the standard identity
\begin{equation}\label{eq:times}
    \int_0^t \text{d}t_s  \int_0^t \text{d}t_s^{\prime} = \int_0^t \text{d}t_s  \int_0^{t_s} \text{d}t_{s}^{\prime} + \int_0^t \text{d}t_s^{\prime} \int_0^{t_s^{\prime}} \text{d}t_s  \, .
\end{equation}
We then find
\begin{align}
    \dot{p}_{\mathcal{R}}(x_1,x_2,t) \simeq &  \int_{0}^{2\pi}\d \varphi_s   \bar{D}_F (x_s ,t | x_1)   \int_0^t \d t_s^\prime \int_{0}^{2\pi}\d \varphi_s^\prime D_F^* ( x_s , t |x_s^\prime , t_s^\prime)  \bar{D}_F^* ( x_s^\prime ,t_s^\prime | x_2 )  \nonumber \\ 
    +&   \int_{0}^{2\pi}\d \varphi_s^\prime   \bar{D}_F^* ( x_s^\prime ,t | x_2 )\int_0^t \d t_s  \int_{0}^{2\pi}\d \varphi_s   D_F ( x_s^\prime , t | x_s , t_s) \bar{D}_F (x_s ,t_s | x_1)  \nonumber \\ 
    =& \int_{0}^{2\pi}\d \varphi_s  \bar{D}_F (x_s ,t | x_1)   D_F^* ( x_s , t |x_2 )  +  \int_{0}^{2\pi}\d \varphi_s \bar{D}_F^* ( x_s ,t | x_2 )  D_F ( x_s , t |x_1) \,  .
\end{align}
We can further decompose the remaining propagators using Eqs.~\eqref{eq:Ddecomp1}-\eqref{eq:Ddecomp3},
\begin{align}
    \dot{p}_{\mathcal{R}}\simeq    \int_{0}^{2\pi}\d \varphi_s &\left[\int_0^{2 \pi} \d \varphi_* \int_0^{t} \d t_* \  \bar{\underaccent{\bar}{D}}_F (x_s,t | x_* ,t_*) D_F(x_*,t_* | x_1)\right] \nonumber \\ \times & \left[\int_{0}^{2\pi}\d \varphi_*^\prime \int_{0}^t \d t_* \underaccent{\bar}{D}_F^*( x_s, t |x_*^\prime, t_* )  D_F^*(x_*^\prime,t_*|x_2)\right]  \nonumber \\ 
    +  \int_{0}^{2\pi}\d \varphi_s & \left[\int_0^{2 \pi} \d \varphi_* \int_0^{t} \d t_* \  \bar{\underaccent{\bar}{D}}_F^* (x_s,t | x_* ,t_*) D_F^*(x_*,t_* | x_2)\right]  \nonumber \\ \times &\left[\int_{0}^{2\pi}\d \varphi_*^\prime \int_{0}^t \d t_* \underaccent{\bar}{D}_F( x_s, t |x_*^\prime, t_* )  D_F(x_*^\prime,t_*|x_1)\right]   \, .
\end{align}
We can now perform the $x_1$- and $x_2$-integrals in Eq.~\eqref{eq:P_R} using another standard identity,
\begin{align}
     \intd x_i \Psi_{E,J}(x_i)  D_F(x_*,t_* | x_i)= \Psi_{E,J}(x_*,t_*)\, . 
\end{align}
This gives
\begin{align}
    \dot{P}_{\mathcal{R}}\simeq  \int_{0}^{2\pi}\d \varphi_s & \left[\int_0^{2 \pi} \d \varphi_* \int_0^{t} \d t_* \  \bar{\underaccent{\bar}{D}}_F (x_s,t | x_* ,t_*) \Psi_{E,J}(x_*,t_*)\right] \nonumber \\ 
    \times & \left[\int_{0}^{2\pi}\d \varphi_*^\prime \int_{0}^t \d t_* \underaccent{\bar}{D}_F^*( x_s, t |x_*^\prime, t_* )  \Psi_{E,J}^*(x_*^\prime,t_*)\right]  \nonumber \\ 
    +  \int_{0}^{2\pi}\d \varphi_s & \left[\int_0^{2 \pi} \d \varphi_* \int_0^{t} \d t_* \  \bar{\underaccent{\bar}{D}}_F^* (x_s,t | x_* ,t_*) \Psi_{E,J}^*(x_*,t_*)\right] \nonumber \\ 
    \times & \left[\int_{0}^{2\pi}\d \varphi_*^\prime \int_{0}^t \d t_* \underaccent{\bar}{D}_F( x_s, t |x_*^\prime, t_* )  \Psi_{E,J}(x_*^\prime,t_*)\right]\, ,
\end{align}
allowing us to identify the second term as the complex conjugate of the first.

The underlying symmetry of the system can now be made explicit by using the form of the wave function, $\Psi_{E,J}(x)=\psi_{E,J}(r) e^{iJ\varphi}$, and introducing the \textit{angular-integrated} propagator 
\begin{align}
    A_{J}(x_f,t|r_i)\equiv&\intd \varphi_i e^{i J \varphi_i} D_F (x_f,t|x_i)\, .
\end{align}
The propagators $\underaccent{\bar}{A}_J$ and $\bar{\underaccent{\bar}{A}}_J$ meanwhile are defined through an analogous relation with $D_F$ replaced by $\underaccent{\bar}{D}_F$ and $\bar{\underaccent{\bar}{D}}_F$, respectively. In terms of these functions, we can express the probability flux as
\begin{align}
    \dot{P}_{\mathcal{R}}\simeq &  \int_{0}^{2\pi}\d \varphi_s \left[ \int_0^{t} \d t_* \  \bar{\underaccent{\bar}{A}}_J (x_s,t | r_* ,t_*) \psi_{E,J}(r_*,t_*)\right] \left[ \int_{0}^t \d t_* \underaccent{\bar}{A}_J^*( x_s, t |r_*^\prime, t_* )  \psi_{E,J}^*(r_*^\prime,t_*)\right] +c.c.  \nonumber \\
    \label{eq:PRdotfull}
    =&\int_{0}^{2\pi}\d \varphi_s \left[ \int_0^{t} \d \Delta t \  \bar{\underaccent{\bar}{A}}_J (x_s, \Delta t | r_*) \psi_{E,J}(r_*,t-\Delta t)\right] \left[ \int_{0}^t \d \Delta t \underaccent{\bar}{A}_J^*( x_s, \Delta t |r_*^\prime)  \psi_{E,J}^*(r_*^\prime,t-\Delta t)\right] +c.c.\,  .
\end{align}
This result is non-perturbative in the sense that it does not directly rely on the semi-classical expansion. We have, however, made several physical assumptions in its derivation. First, in order to define the radii $r_*$ and $r_s$ (or, equivalently, the submanifolds $\Sigma_*$ and $\Sigma_s$), we have assumed that the wave function is sharply localised within a region $r<r_*$. For our example of a resonance state, this localisation can be deduced from the WKB form of the wave function~\eqref{eq:WKBres}. Second, in Eq.~\eqref{eq:nobacktunnel}, we had assumed the absence of back-tunnelling. For a bounded potential, this assumption would be violated once a significant fraction of the probability distribution has been transferred from $\mathcal{F}$ to $\mathcal{R}$, implying an upper bound on the physical time~\cite{Andreassen:2016cff}. At the same time, we have made the fourth assumption that the resonance state acts as an approximate eigenstate of the Hamiltonian. This assumption is also only justified for relatively small physical times before it is violated by non-linear effects~\cite{Lin:2025bjn}. Together, this implies that Eq.~\eqref{eq:PRdotfull} is only valid for $t<t_{\rm NL}$~\cite{Andreassen:2016cff}. Our fifth assumption on the other hand requires that $t\gg t_{\rm sys}$, as necessary for the steadyon to approximately satisfy the relevant boundary conditions~\cite{Steingasser:2024ikl}. Thus, the applicability of Eq.~\eqref{eq:PRdotfull} can be summarised through the condition $t_{\rm sys}<t<t_{\rm NL}$, with the existence of such a regime posing a non-trivial condition in itself. The latter is, however, generally satisfied in the semi-classical limit invoked in the next section.

\section{Semi-classical evaluation}\label{sec:semi-class}

In this section, we discuss the implications of Eq.~\eqref{eq:PRdotfull} in the semi-classical limit. In Sec.~\ref{sec:SCRealTime}, we evaluate the tunnelling rate to leading order using the steadyon framework. Then, in Sec.~\ref{sec:euclidean_picture} we show that our results can be rephrased in terms of Euclidean-time quantities, thereby significantly simplifying calculations. For the readers' convenience, we summarise our results together with step-by-step instructions for the calculation of the tunnelling rate in Sec.~\ref{sec:recipe}. Finally, we provide a detailed discussion of a full example in Sec.~\ref{sec:example}.

\subsection{Real-time calculation}\label{sec:SCRealTime}

In Eq.~\eqref{eq:PRdotfull}, we have found for the probability flux controlling the tunnelling rate
\begin{align}
    \dot{P}_{\mathcal{R}}\simeq & \int_{0}^{2\pi}\d \varphi_s \left[ \int_0^{t} \d \Delta t \  \bar{\underaccent{\bar}{A}}_J (x_s, \Delta t | r_*) \psi_{E,J}(r_*,t-\Delta t)\right] \left[ \int_{0}^t \d \Delta t \underaccent{\bar}{A}_J^*( x_s, \Delta t |r_*^\prime)  \psi_{E,J}^*(r_*^\prime,t-\Delta t)\right] +c.c. \, ,
\end{align}
where 
\begin{align}
    A_J(x_s,\Delta t|r_*)  =  \intd \varphi_* e^{i J \varphi_*} D_F (x_s,\Delta t|x_*)=  \intd \varphi_* \ e^{i J \varphi_*} \int_{x(0)=x_*}^{x(\Delta t)=x_s} \mathcal{D}x\ e^{i S[x]}\, ,\label{eq:Acomplete}
\end{align}
and similarly for $\bar{A}_J$ and $\bar{\underaccent{\bar}{A}}_J$. The time-dependence of $\psi_{E,J}(x_*)$, meanwhile, is determined by our initial assumption of a resonance state, 
\begin{align}
\label{eq:time_dependence}
    \psi_{E,J}(r_*,t)\simeq \psi_{E,J}(r_*,0)\cdot e^{-i E t}\, .
\end{align}
The real-time path integral in Eq.~\eqref{eq:Acomplete} can now be evaluated using the \textit{steadyon picture} developed in Refs.~\cite{Steingasser:2024ikl,Steingasser:2024wkh,Lin:2025pap,Lin:2025bjn}. This technique is based on a regularisation of the theory through the introduction of an infinitesimal imaginary part for the Hamiltonian,
\begin{align}\label{eq:Reg}
    H \to (1-i \epsilon)H\, ,
\end{align}
with $\epsilon$ being some arbitrary but small constant. This regularisation then also manifests in a complexification of the action,
\begin{align}
    S_\epsilon=(1+i\epsilon) \int_0^{\Delta t} \d t \ \frac{m}{2}\left( \dot{r}^2 + r^2 \dot{\varphi}^2 \right) - (1 - 2i \epsilon) V(r)\, .
\end{align}
The complexification also affects the time evolution of the wave function, which is now given by~\cite{Lin:2025bjn}
\begin{align}
    \psi_{E,J}(r_*,t-\Delta t)\simeq \psi_{E,J}(r_*,0)\cdot e^{-i E (t-\Delta t) - \epsilon E (t-\Delta t)}\, .
\end{align}
Thus, in this regularised theory each of the factors in Eq.~\eqref{eq:PRdotfull} takes the form
\begin{align}
    F\equiv & \int_0^{t} \d \Delta t \  \intd \varphi_* \  \int_{x(0)=x_*}^{x(\Delta t)=x_s} \mathcal{D}x\ e^{i S_\epsilon[x]+ i J \varphi_*-i E (t-\Delta t) - \epsilon E (t-\Delta t)}  \psi_{E,J}(r_*,0)\nonumber \\ 
    \label{eq:factor_F}
    =&  e^{-i E t - \epsilon E t} \int_0^{t} \d \Delta t \  \intd \varphi_* \  \int_{x(0)=x_*}^{x(\Delta t)=x_s} \mathcal{D}x\ e^{i S_\epsilon[x]+ i J \varphi_* +i (1-i \epsilon) E \Delta t  }  \psi_{E,J}(r_*,0)\, .
\end{align}
To leading order, this expression can be evaluated in the semi-classical regime through a saddle point approximation for the three integrals.\footnote{For notational simplicity, we do not distinguish between stationary phase and saddle point approximations throughout this article.} In polar coordinates, extremising the exponent with respect to the particle's trajectory, initial angle, and $\Delta t$ amounts to the conditions
\begin{align}
    0= & \frac{\delta S [\bar{r},\bar{\varphi}]}{\delta \bar{r}(t)}= m \ddot{\bar{r}}(t) - m \bar{r}(t) \dot{\bar{\varphi}}^2(t) + (1 - 2i \epsilon)V^\prime (\bar{r}(t)) \ , \label{eq:SPrbar} \\ 
    0=&\frac{\delta  S [\bar{r},\bar{\varphi}]}{\delta \bar{\varphi}(t)}= 2 \dot{\bar{r}}(t)  \dot{\bar{\varphi}}(t) + \bar{r}(t) \ddot{\bar{\varphi}}(t) \ , \label{eq:SPphibar} \\
    0=& \frac{d}{d \varphi_*}  S_\epsilon [\bar{r},\bar{\varphi}]+J \ , \label{eq:SPphis}\\
    0=& \frac{d}{d \Delta t}S_{\epsilon} [\bar{r},\bar{\varphi}]+(1-i\epsilon) E\, ,\label{eq:SPdT}
\end{align}
with the \textit{steadyon} $\{ \bar{r}(t),\bar{\varphi}(t)\}$. Note that, in addition to these relations, the steadyon is also subject to the crossing conditions we introduced around Eq.~\eqref{eq:Dbar}--\eqref{eq:Dbarunderbar}.

We can now first integrate Eq.~\eqref{eq:SPphibar}, which gives the usual conservation of angular momentum,
\begin{align}
\label{eq:J_conservation}
    \bar{J}=m \bar{r}^2(t)\dot{\bar{\varphi}}(t)={\rm const.}\, ,
\end{align}
with some yet to be determined constant $\bar{J}$. Using this relation, Eq.~\eqref{eq:SPrbar} can be converted to a closed equation for $\bar{r}(t)$,
\begin{align}
    0= &  m \ddot{\bar{r}}(t) - m \bar{r}(t) \dot{\bar{\varphi}}^2(t) + (1 - 2i \epsilon)V^\prime (\bar{r}(t)) = m \ddot{\bar{r}}(t) - \frac{\bar{J}^2}{m \bar{r}^3(t)} + (1 - 2i \epsilon)V^\prime (\bar{r}(t)) \nonumber \\ 
    =&  m \ddot{\bar{r}}(t) + (1 - 2i \epsilon)V_{\rm eff}^\prime (\bar{r}(t))\, , \label{eq:rsteady}
\end{align}
where
\begin{align}
    V_{\rm eff}(r)=V(r) + \frac{(1 + 2i \epsilon)\bar{J}^2}{2m r^2}\, .
\end{align}

To determine the remaining integration constant $\bar{J}$, we can now evaluate Eq.~\eqref{eq:SPphis}. As a first step, we consider a steadyon $\{\bar{r},\bar{\varphi}\}$ with a fixed final and initial points $x_s$ and $x_*$, respectively. We then introduce an infinitesimal perturbation in the initial angle, $\varphi_* \to \varphi_* + \delta \varphi_*$, which in turn induces an infinitesimal perturbation in the steadyon,
\begin{align}
    \{\bar{r},\bar{\varphi}\} \to \{\bar{r} + \delta r,\bar{\varphi}+\delta \varphi\}\, .
\end{align}
This then causes a change in the action,
\begin{align}
    S_\epsilon[\bar{r} + \delta r,\bar{\varphi}+\delta \varphi] \simeq & S_\epsilon[\bar{r},\bar{\varphi}] + \int \left[ \frac{\partial L}{\partial r} \delta r + \frac{\partial L}{\partial \dot{r}} \delta \dot{r} +  \frac{\partial L}{\partial \varphi} \delta \varphi + \frac{\partial L}{\partial \dot{\varphi}} \delta \dot{\varphi} \right] \d t \nonumber \\ 
    =& S_\epsilon[\bar{r},\bar{\varphi}] + \int  \left[\left( \frac{\partial L}{\partial r} - \frac{\d}{\d t} \frac{\partial L}{\partial \dot{r}} \right) \delta r + \left(  \frac{\partial L}{\partial \varphi} -\frac{\d}{\d t}  \frac{\partial L}{\partial \dot{\varphi}} \right) \delta \varphi \right] \d t + \frac{\partial L}{\partial \dot{r}} \delta r\bigg|_{r_*}^{r_s} + \frac{\partial L}{\partial \dot{\varphi}} \delta \varphi\bigg|_{\varphi_*}^{\varphi_s} \nonumber \\ 
    =& \left[(1+i \epsilon)\left(m \bar{r}^2 \dot{\bar{\varphi}} \right) \delta \varphi \right]_{\varphi_*}^{\varphi_s} = - (1+i \epsilon) \bar{J} \cdot \delta \varphi_* \label{eq:dSvarphi}\, .
\end{align}
This now allows us to relate the steadyon's angular momentum to that of the initial state:
\begin{align}
    \frac{\d S_\epsilon}{\d \varphi_*}=- (1+ i \epsilon) \bar{J} \quad \overset{\text{\eqref{eq:SPphis}}}{\Rightarrow}\quad (1+ i \epsilon) \bar{J} = J\quad  \overset{\text{\eqref{eq:J_conservation}}}{\Rightarrow}\quad \dot{\bar{\varphi}}(t) = \frac{(1-i \epsilon)J}{m \bar{r}^2(t)}\, . \label{eq:Jini}
\end{align}
Thus, we find for the action of the steadyon
\begin{align}
    S_\epsilon[\bar{r},\bar{\varphi}]=&(1+i\epsilon) \int_0^{\Delta t} \d t \ \frac{m}{2}\left( \dot{\bar{r}}^2 + \bar{r}^2 \dot{\bar{\varphi}}^2 \right) - (1 - 2i \epsilon) V(\bar{r})\nonumber \\ 
    =& (1+i\epsilon) \int_0^{\Delta t} \d t \ \frac{m}{2}\dot{\bar{r}}^2 + (1-2i \epsilon)\left[ \frac{ J^2}{2 m \bar{r}^2} -  V(\bar{r})\right]\, , \label{eq:Sepssteady} 
\end{align}
where $\bar{r}(t)$ is the solution of Eq.~\eqref{eq:rsteady}.

To evaluate Eq.~\eqref{eq:SPdT}, we first recall that the initial velocity in angular direction is fixed by Eq.~\eqref{eq:Jini},
\begin{align}\label{eq:phidotin}
    \dot{\bar{\varphi}}(0)=(1-i\epsilon)\frac{J}{m r_*^2}\, .
\end{align}
Next, we can evaluate Eq.~\eqref{eq:SPdT} using the results of Ref.~\cite{Lin:2025pap}, which imply that Eq.~\eqref{eq:SPdT} amounts to setting the steadyon's initial velocity through
\begin{gather}
    (1-2i \epsilon)[E-V(r_*)]=\frac{m}{2}\dot{\bar{x}}^2(0)=\frac{m}{2}\dot{\bar{r}}^2(0)+ \frac{m}{2}\bar{r}^2(0)\dot{\bar{\varphi}}^2(0)= \frac{m}{2}\dot{\bar{r}}^2(0) + \frac{(1-2 i \epsilon)J^2}{2m r_*^2}\nonumber \\ 
    \Rightarrow \frac{m}{2}\dot{\bar{r}}^2(0)=(1-2i \epsilon)[E-V_{\rm eff}(r_*)]=0\, .
\end{gather}
Together, these equations unambiguously determine the relevant steadyon and thus its action.

So far, we have focused on evaluating the modified propagator $A_J$ together with its associated integral over $\Delta t$. To understand the dependence of our result on $\varphi_s$, as necessary to perform the remaining integral, we first observe that the saddle point equation for $\varphi_*$ was equivalent to setting the angular momentum of the steadyon, whose conservation followed from the equation for the steadyon's corresponding component $\bar{\varphi}$. Together, these imply that $\varphi_*$ depends on $\varphi_s$. This can be made manifest through
\begin{align}\label{eq:varphistarofs}
    \bar{\varphi}_*=-(\varphi_s -\bar{\varphi}_*)+  \varphi_s= - \int_0^{\Delta t_{\rm per}} \dot{\bar{\varphi}} \ \d t+  \varphi_s= -\int_0^{\Delta t_{\rm per}}  \frac{\bar{J}}{m \bar{r}^2} \d t +  \varphi_s\, ,
\end{align}
where $\bar{J}=(1-i \epsilon)J$, as in Eq.~\eqref{eq:J_conservation}. The system's underlying symmetry meanwhile implies that the action $S_\epsilon$, as well as the integral term in Eq.~\eqref{eq:varphistarofs}, are invariant under a shift in $\varphi_s$, as the latter would be compensated for by a shift in $\varphi_*$.

As $\bar{r}$ is complex, Eq.~\eqref{eq:varphistarofs} implies that a real $\bar{\varphi}_*$ generally corresponds to a complex $\varphi_s$ and conversely. And indeed, the periodic features of the steadyon together with the symmetry underlying the problem suggest that we are free to shift $\varphi$ by a constant in such a way that either of the two variable is real. We confirm this explicitly for a concrete example in Sec.~\ref{sec:ex}. As our arguments so far did not rely on any assumption on $\varphi_*$, we may select $\varphi_s$ to be real, reducing the $\varphi_s$-dependence to a phase. This allows us to further simplify our expression for $F$, as given in Eq.~\eqref{eq:factor_F}, as
\begin{align}
     F=&\int_0^{t} \d \Delta t \ A_J (x_s, \Delta t | r_*) \psi_{E,J}(r_*,t-\Delta t)\sim  e^{-i E t - \epsilon E t} e^{iS_\epsilon[\bar{r},\bar{\varphi}]+ i J \bar{\varphi}_* +i (1-i \epsilon) E \Delta t_{\rm per}}  \psi_{E,J}(r_*,0)  \nonumber \\ 
     =&  e^{-i E (t- \Delta t_{\rm per}) +i J \varphi_s }e^{ - \epsilon E t + E \epsilon \Delta t_{\rm per}} \exp\left[i \left(S_\epsilon[\bar{r},\bar{\varphi}]- (1-i \epsilon)\int_0^{\Delta t_{\rm per}} \frac{J^2}{m \bar{r}^2} \d t \right) \right]  \psi_{E,J}(r_*,0)\, .
\end{align}
This expression can now be brought to a more compact form in terms of the \textit{Routhian action} $S_{R,\epsilon}$.\footnote{The integrand of the Routhian action, dubbed the \textit{Routhian}, is defined as the partial Legendre transform of the Lagrange function with respect to a cyclic variable. For more detailed discussions, see, e.g., Refs.~\cite{Routh1860,Goldstein2001,LandauLifshitz1976}.} We define this quantity for the regularised theory through
\begin{align}
    S_{R,\epsilon}[r;J]\equiv\int_0^{\Delta t}\d t\ R_\epsilon(r,\dot{r},J)\equiv S_\epsilon[r,\bar{\varphi}(r,\bar{J})] - \int_0^{\Delta t}\d t\ J \dot{\bar{\varphi}}(r,\bar{J})\, ,\label{eq:Sredsteady}
\end{align}
with $\bar{\varphi}(r,J)$ as in Eq.~\eqref{eq:Jini}. For our mechanical system, this transformation amounts to transitioning to an effective action for the radial variable constructed using the effective potential,
\begin{align}\label{eq:SRouth}
    S_{R,\epsilon}[r;J]=(1+i \epsilon)\int_0^{\Delta t}\d t\ \frac{m}{2}\dot{r}^2 - (1-2i\epsilon) V_{\rm eff}(r), \quad {\rm where} \quad V_{\rm eff}(r)= V(r)+ \frac{J^2}{2m r^2}\, .
\end{align}
Just as for the purely classical theory, the effective equation for $\bar{r}$ can be obtained through the variation of this functional, rather than the original action $S_\epsilon$.

In the overall expression for $\dot{P}_{\mathcal{R}}$ the phases cancel, leading to the more compact expression
\begin{align}
\label{eq:P_R_final2}
    \dot{P}_{\mathcal{R}}\sim  e^{ -2 \epsilon t E} e^{-2 {\rm Im}(S_{R,\epsilon}[\bar{r}]) +2 E  \epsilon \Delta t_{\rm per}} |\psi_{E,J}(r_*,0)|^2\, .
\end{align}
To evaluate the expression for $\Gamma$, it now only remains to calculate the denominator $P_{\mathcal{F}}(t)$. Taking into account the deformed Hamiltonian and the resulting time-dependence in Eq.~\eqref{eq:time_dependence}, we find
\begin{align}
    P_{\mathcal{F}}(t)=\int_{\mathcal{F}} \d^2x|\Psi_{E,J}(x,t)|^2=e^{-2\epsilon t E}\int_{\mathcal{F}} \d^2x|\Psi_{E,J}(x,0)|^2\sim e^{-2\epsilon t E}\, .
\end{align}
The denominator therefore cancels precisely with the first factor in Eq.~\eqref{eq:P_R_final2}, removing the dependence on physical time and leading us to the final leading-order expression
\begin{align}
\label{eq:Gamma_final}
    \Gamma\sim  e^{-2 {\rm Im}(S_{R,\epsilon}[\bar{x}]) +2 E  \epsilon \Delta t_{\rm per}} |\psi_{E,J}(r_*,0)|^2\, .
\end{align}

\subsection{Euclidean-time picture}
\label{sec:euclidean_picture}
\begin{figure}[t!]
\centering
\vspace*{-0.5cm}
\begin{tikzpicture}[scale=2.5,
    line width=1.3pt,
    font=\Large]
    \draw[->](0,0)--(4.4,0); 
    \draw[->](0,0)--(0,-1.9); 
    \draw[black, thick](0,0.03)--(0,-0.03);
    \draw (-0.13,0.13) node [black]{$0$};
    \draw (4.3,-0.16) node [black]{$t$};
    \draw (4.6,0) node [black]{$\Delta t$};
    \draw[black, thick](4.3,0.03)--(4.3,-0.03);
    \draw (0,-2.1) node [black]{$\Delta \tau$};
    \draw (0.475,-0.09) node [black] {$\epsilon$};
    \draw [gray!60,ultra thick](0.74,0) arc [start angle=0, end angle=-60, radius=0.3]; 
    \draw[gray, dashed](0,-1.4894)--(3.5,-1.4894);
    \draw[gray, thick](3.5,0.03)--(3.5,-0.03);
    \draw[gray, thick](-0.03,-1.4894)--(0.03,-1.4894);
    \draw (-0.4,-1.4894) node [black]{$\Delta \tau_{\rm per}$}; 
    \draw (3.5,0.3) node [black]{$\Delta t_{\rm per} $}; 
    \draw[oiBlue,very thick,->](0,0)--(0.5,0); 
    \draw[oiBlue,very thick,->](0,0)--(1,0); 
    \draw[oiBlue,very thick,->](0,0)--(1.5,0); 
    \draw[oiBlue,very thick,->](0,0)--(2,0); 
    \draw[oiBlue,very thick,->](0,0)--(2.5,0); 
    \draw[oiBlue,very thick,->](0,0)--(3,0); 
    \draw[oiBlue,very thick,->](0,0)--(3.5,0); 
    \draw (1.75,0.3) node [text=oiBlue]{$\gamma_t$}; 
    \draw[oiOrange,very thick,->](3.5,0)--(3.5,-1.4894);  
    \draw[oiOrange, very thick,->](3.5,0)--(3.5,-0.496466); 
    \draw[oiOrange, very thick,->](3.5,0)--(3.5,-0.992933); 
    \draw (3.8,-0.75) node [text=oiOrange]{$\gamma_\tau$}; 
    \draw[oiGreen, very thick,->](0,0)--(3.5,-1.4894); 
    \draw[oiGreen, very thick,->](0,0)--(0.7,-0.29788); 
    \draw[oiGreen, very thick,->](0,0)--(1.4,-0.59576);
    \draw[oiGreen, very thick,->](0,0)--(2.1,-0.89364);
    \draw[oiGreen, very thick,->](0,0)--(2.8,-1.19152);
    \draw (1.5,-0.9) node [text=oiGreen]{$\gamma_\epsilon$}; 
\end{tikzpicture}
\caption{The complex-time plane with the three contours of interest for our argument. For a system without explicit time-dependence, the regularisation of the Hamiltonian underlying the steadyon picture can equivalently be interpreted as an infinitesimal Wick rotation leading to the diagonal contour $\gamma_\epsilon$ (green). For a given $\epsilon$, the total (unphysical) tunnelling time $\Delta t_{\rm per}$ is determined by extremising the $\Delta t$-integral in Eq.~\eqref{eq:factor_F}. This contour can be deformed into a segment along the real-time axis, $\gamma_t$ (blue) and one along the Euclidean-time axis, $\gamma_\tau$ (orange).}
\label{fig:sketck}
\end{figure}
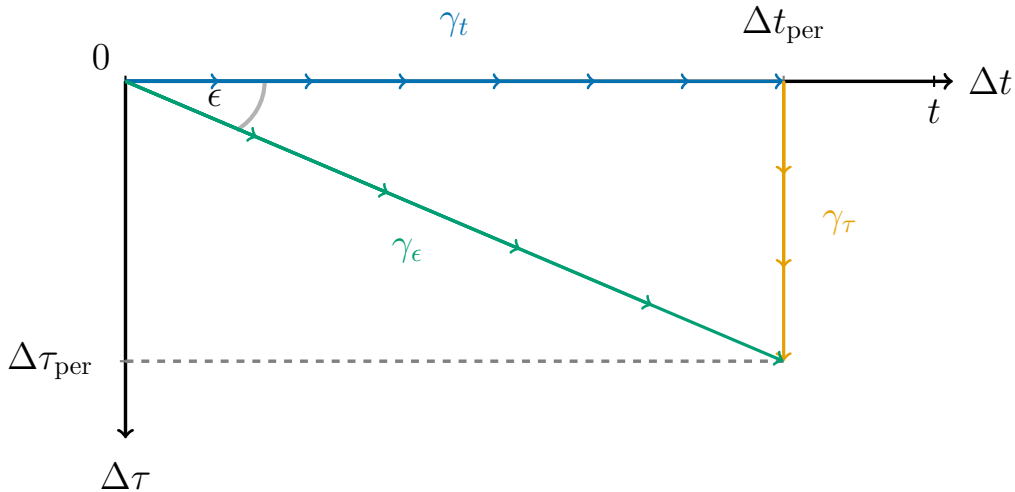

\noindent
One of the most important features of the steadyon picture is its ability to yield more easily solvable Euclidean-time equations in the relevant limits of a vanishing regulator, $\epsilon \ll 1$, and a large tunnelling time relative to the system's typical time scale, $t \gg t_{\rm sys}$. This connection can be made transparent by observing that the regularisation~\eqref{eq:Reg} can equivalently be assigned to the time variable rather than the Hamiltonian,
\begin{align}
    U(t)=\exp (-i Ht)\to \exp (-i \underbrace{H (1-i \epsilon)}_{\rm regularisation}\cdot\, t)= \exp (-i H\cdot \hspace{-2mm} \underbrace{(1-i \epsilon)t}_{\rm Wick\ rotation} \hspace{-2mm})\, .
\end{align}
Thus, the regularisation can equivalently be understood as an infinitesimal Wick rotation, $t\to e^{-i \epsilon} t$, as shown in Fig.~\ref{fig:sketck}.

This interpretation is also consistent with the deformed action and resulting equations of motion, which can similarly be understood as those of the usual theory analytically extended onto the diagonal contour $\gamma_\epsilon$ with $z=e^{-i \epsilon}t$. Thus, the steadyon can be analytically extended to the complex-time plane. In the following, we will show how this observation can be used to map the tunnelling rate onto a familiar, and computationally simpler, Euclidean-time problem. To do so, we will consider two complex-time contours besides $\gamma_\epsilon$, as depicted in Fig.~\ref{fig:sketck}.

First, we consider a contour $\gamma_t$, which can be parametrised through $z(t)=t$ with $t\in [0, \Delta t_{\rm per}]$, which implies that furthermore $\d z=\d t$ and $\d /\d z=\d /\d t$. We will highlight the projection of the steadyon onto this contour with an index ``cl'', i.e.,
\begin{align}
    \bar{r}_{\rm cl}(t)\equiv \bar{r}(z=t) \quad {\rm and} \quad \bar{\varphi}_{\rm cl}(t)\equiv \bar{\varphi}(z=t)\, .
\end{align}
Next, we consider a contour parallel to the Euclidean-time axis $\gamma_\tau$, which we parametrise through $z(\tau)=\Delta t_{\rm per}-i \tau$, with $\tau \in [0, \Delta \tau_{\rm per}]$, where $\Delta \tau_{\rm per}\simeq \epsilon \cdot \Delta t$. Along this contour, $\d z =-i \d \tau$ and $\frac{\d}{\d z}=i \frac{\d}{\d \tau}$. In anticipation of our later results, we will denote the projection of the steadyon onto this contour with a label $I$,
\begin{align}
    \bar{r}_I(\tau)\equiv \bar{r}(z=\Delta t_{\rm per}-i \tau) \quad {\rm and} \quad \bar{\varphi}_I(\tau)\equiv \bar{\varphi}(z=\Delta t_{\rm per}-i \tau)\, .
\end{align}

To arrive at the Euclidean-time picture, we will now consider two equivalent approaches: first, we will consider the effective theory obtained by integrating out the angular motion using the conservation of angular momentum; second, discuss the full two-dimensional model, with a particular focus on the properties of the full instanton solution.

\subsubsection{Effective one-dimensional description}

In Eq.~\eqref{eq:Gamma_final}, we found that the tunnelling rate is, to leading order, controlled by the Routhian action, whose variation yields the equation of motion for the radial component of steadyon. Starting from Eq.~\eqref{eq:SRouth} it is straightforward to identify the regularised Routhian action with the integral of the corresponding Routhian along $\gamma_\epsilon$,
\begin{gather}
\label{eq:action_z}
    S_{R,\epsilon}[r;J]= \int_{\gamma_\epsilon} \d z \ \frac{m}{2}\left(\frac{\d r}{\d z}\right)^2  -  V_{\rm eff}(r(z))\, .
\end{gather}
The equation for $\bar{r}$ and its boundary conditions can be similarly recast in terms of a complex-time variable $z$,
\begin{align}
\label{eq:eom_z}
    m\frac{\d^2 \bar{r}}{\d z^2}= - V^\prime_{\rm eff}(\bar{r}), \quad \bar{r}(0)=r_*, \quad \dot{\bar{r}}(0)=0, \quad \bar{r}(\Delta z)=r_s, \quad \dot{\bar{r}}(\Delta z)=0\, ,
\end{align}
with $\Delta z=(1-i\epsilon )\Delta t$.\footnote{Recall that the steadyon, as constructed in Sec.~\ref{sec:FPderivation}, generically only satisfies the boundary condition at $\Delta z$ up to an offset of $\mathcal{O}(\epsilon)$. We moreover found the condition $\Delta t_{\rm per}\cdot \epsilon \simeq \Delta \tau_{\rm per}$. Thus, these conditions are only satisfied in the combined limit $\epsilon \ll 1$ and $t > \Delta t_{\rm per} \gtrsim \Delta \tau_{\rm per}/\epsilon$.} This allows for an immediate projection of the steadyon onto the two contours parallel to the real and Euclidean time axes, respectively, as well as the evaluation of the Routhian action along these contours.

Along $\gamma_t$, Eq.~\eqref{eq:eom_z} reproduces the classical equation of motion for $\bar{r}_{\rm cl}(t)$. Thus, along this contour, the particle undergoes its classical dynamics, while the projection of the action~\eqref{eq:action_z} onto $\gamma_t$ coincides with its classical action. This implies, in particular, that ${\rm Im}(S_{\gamma_t})=0$, corresponding to a vanishing contribution to the tunnelling rate.
 
To understand the dynamics along $\gamma_\tau$, we first note that $\Delta t_{\rm per}$ is defined by requiring that $\bar{r}_{\rm cl}(0)=r_*=\bar{r}_{\rm cl}(\Delta t_{\rm per})$, up to a correction controlled by $\epsilon$, which may be neglected in the limit $\epsilon \to 0$. Thus, we find that the projection of the steadyon onto $\gamma_\tau$, $\bar{r}_I(\tau)$, is subject to the equation of motion and boundary conditions
\begin{align}
    m\frac{\d ^2 \bar{r}_I}{\d \tau^2}=V^\prime_{\rm eff}(\bar{r}_I), \quad \bar{r}(0)=r_*, \quad \bar{r}(\Delta \tau_{\rm per})=r_s, \quad  \frac{\d \bar{r}}{\d \tau}(0)=0\, .
\end{align}
\begin{figure}
    \centering
    \includegraphics[width=\linewidth]{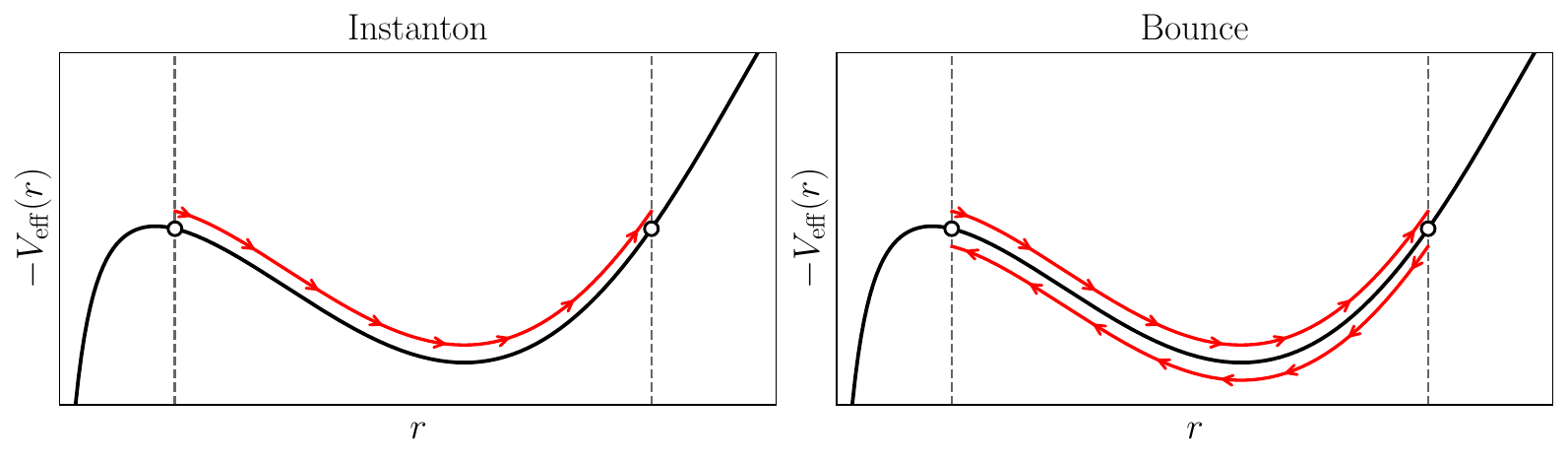}
    \caption{Euclidean radial motion in the inverted effective potential \(-V_{\rm eff}(r)\). \textit{Left}: the instanton branch crossing once from \(r_\ast\) to \(r_s\) in Euclidean time \(\Delta\tau_{\rm per}\). \textit{Right}: the corresponding periodic bounce obtained by continuing the motion back to \(r_\ast\), with total period \(2\Delta\tau_{\rm per}\).}
    \label{fig:instanton}
\end{figure}
We can now immediately recognise this as an equation of motion and boundary conditions of a familiar (periodic) instanton in the potential $V_{\rm eff}$, while the crossing condition allows us to identify $\Delta \tau_{\rm per}$ with the Euclidean-time interval necessary for the instanton to move exactly once from $r_*$ to $r_s$. For the purpose of efficiently calculating the tunnelling rate, this relation can be used as a definition of $\Delta \tau_{\rm per}$ directly within the Euclidean-time picture. See Fig.~\ref{fig:instanton}. The complex Routhian action of this solution along $\gamma_\tau$ meanwhile can be identified with the strictly Euclidean action of this instanton up to a factor of $i$ from the line element. Thus,
\begin{align}
    S_{R,\gamma_\tau}[\bar{r}_I;J]
    &= \int_{\gamma_\tau} \d z \left[ \frac{m}{2}\left(\frac{\d \bar{r}_I}{\d z}\right)^2  - V_{\rm eff}(\bar{r}_I)\right] = \int_0^{\Delta \tau_{\rm per}} (-i\,\d\tau)
    \left[
    \frac{m}{2}\left(i \frac{\d \bar{r}_I}{\d \tau}\right)^2
    -V_{\rm eff}(\bar{r}_I)
    \right] \nonumber\\
    &= i \int_0^{\Delta \tau_{\rm per}} \d\tau
    \left[
    \frac{m}{2} \left(\frac{\d \bar{r}}{\d \tau}\right)^2
    + V_{\rm eff}(\bar{r}_I)
    \right] = i S_{R,E} [\bar{r}_I]\, .
    \label{eq:S_vertical_raw_new}
\end{align}
To relate to the existing literature, we can now also introduce the \textit{bounce} solution $\bar{r}_b(\tau)$ as the instanton followed by the particle rolling back from $r_s$ to $r_*$, as shown in Fig.~\ref{fig:instanton}. Together with Eq.~\eqref{eq:Gamma_final}, this allows us to rewrite the exponent of the tunnelling rate as
\begin{align}
    \ln \Gamma\simeq - 2 {\rm Im} (S_{R,\epsilon}[\bar{r}]) + 2 E \Delta \tau_{\rm per}=-2  S_{R,E}[\bar{r}_I] +2  E \Delta \tau_{\rm per}=- S_{R,E}[\bar{r}_b] +2 E \Delta \tau_{\rm per}\, .
\end{align}
This one-dimensional description is, in fact, sufficient to understand the full two-dimensional description, also taking into account the angular coordinate. This can be made manifest by considering the full Euclidean action,
\begin{align}
    S_{E}[r]=\intd \tau\  \frac{m}{2}\left(\frac{\d r}{\d \tau}\right)^2 +  \frac{m}{2} r^2\left(\frac{\d \varphi}{\d \tau}\right)^2 + V(r)\, .
\end{align}
This action implies an analogous relation for the conserved angular momentum as in real time,
\begin{align}
    m \bar{r}^2(\tau)\frac{\d\bar{\varphi}}{\d \tau}(\tau)=J_E={\rm const.}\, ,
\end{align}
which in turn gives rise to a potential barrier of the form
\begin{align}
    V_{\rm eff}(r)= V(r)-\frac{J_E^2}{2 m r^2}\, .
\end{align}
This, however, is only consistent with the result obtained from the projection of the steadyon if $J_E=\pm i J$.

\subsubsection{Two-dimensional description}

To understand the transition from real to Euclidean time in the full theory, we observe that the regularised action can also be identified with an integral of an analytic function along $\gamma_\epsilon$,
\begin{align}
    S_{\gamma_\epsilon}[r] = \int_{\gamma_\epsilon} \d z \left[ \frac{m}{2}\left(\frac{\d r}{\d z}\right)^2 + \frac{m}{2} r^2 \left(\frac{\d \varphi}{\d z}\right)^2 - V(r)\right]\, .
    \label{eq:Sz_full2d_new}
\end{align}
It is again straightforward to obtain the analytic continuation of this expression, whose corresponding equations of motion can be identified with the continuation of the steadyon equations,
\begin{gather}
\label{eq:eom_r_full2d_new}
    m \frac{\d^2 \bar{r}}{\d z^2} - m \bar{r} \left(\frac{\d \bar{\varphi}}{\d z}\right)^2 + V'(\bar{r}) = 0, \\
    \label{eq:eom_phi_full2d_new}
    \frac{\d}{\d z} \left( m \bar{r}^2 \frac{\d \bar{\varphi}}{\d z} \right) = 0\, .
\end{gather}
The second of these equations implies a constant angular momentum throughout the complex-time plane, thus fully determining $\bar{\varphi} (z)$ as a function of $\bar{r}$ up to a constant. The latter is again determined by the angular momentum of the initial state through Eq.~\eqref{eq:phidotin},
\begin{align}
    \frac{\d \bar{\varphi}}{\d z}(0)=\frac{J}{m r_*^2}\, .
\end{align}
Along the real- and Euclidean-time contours, $\gamma_t$ and $\gamma_\tau$, respectively, this implies
\begin{align}
     z = t \Rightarrow & \qquad \frac{\d \bar{\varphi}_{\rm cl}}{\d t}(0)=\frac{J}{m r_*^2},\\
     z = -i \tau \Rightarrow & \qquad \frac{\d \bar{\varphi}_I}{\d \tau}(0)=-i\frac{J}{m r_*^2} \overset{!}{=} \frac{J_E}{m r_*^2}\, .\label{eq:JEucl}
\end{align}
The Routhian action controlling the tunnelling rate, meanwhile, is 
\begin{align}
    S_{R,\gamma_\epsilon}[\bar{r};J] = &\int_{\gamma_\epsilon} \d z  \frac{m}{2}\left(\frac{\d \bar{r}}{\d z}\right)^2 + \frac{m}{2} \bar{r}^2 \left(\frac{\d \bar{\varphi}}{\d z}\right)^2 - V(\bar{r}) - J \frac{\d \bar{\varphi}}{\d z}\, .
    \label{eq:SRz_full2d_new}
\end{align}
As for the previous case, we first consider the projection of the steadyon onto $\gamma_t$. Along this contour, Eqs.~\eqref{eq:eom_r_full2d_new}-\eqref{eq:eom_phi_full2d_new} again reproduce the classical dynamics, thus leading to a strictly real action, ${\rm Im}(S_{\gamma_t})=0$.

Along the Euclidean-time contour $\gamma_\tau$, Eqs.~\eqref{eq:eom_r_full2d_new}--\eqref{eq:eom_phi_full2d_new} simply reproduce the Euclidean-time equations of motion,
\begin{gather}
    m \frac{\d^2 \bar{r}_I}{\d \tau^2} - m \bar{r}_I \left(\frac{\d \bar{\varphi}_I}{\d \tau}\right)^2 - V'(\bar{r}_I) = 0, \label{eq:eom_r_Eucl_full2d_new}\\
    \frac{\d}{\d\tau}\left(m \bar{r}_I^2 \frac{\d \bar{\varphi}_I}{\d \tau}\right) = 0 \overset{\text{\eqref{eq:JEucl}}}{\Longrightarrow} m \bar{r}_I^2 \frac{\d \bar{\varphi}_I}{\d \tau}= J_E=-i J\, .\label{eq:eom_phi_Eucl_full2d_new}
\end{gather}
The Routhian action along this contour similarly reproduces the Euclidean action up to a factor of $i$,
\begin{align}
    S_{R,\gamma_\tau}[\bar{r}_I;J]
    &= \int_{\gamma_\tau} \d z \left[ \frac{m}{2}\left(\frac{\d \bar{r}_I}{\d z}\right)^2 + \frac{m}{2} \bar{r}_I^2 \left(\frac{\d \bar{\varphi}_I}{\d z}\right)^2 - V(\bar{r}_I) - J \frac{\d \bar{\varphi}}{\d z}\right] \nonumber\\
    &= \int_0^{\Delta \tau_{\rm per}} (-i\,\d\tau)
    \left[
    \frac{m}{2}\left(i \frac{\d \bar{r}_I}{\d \tau}\right)^2
    + \frac{m}{2} \bar{r}_I^2 \left(i \frac{\d \bar{\varphi}_I}{\d \tau} \right)^2
    - V(\bar{r}_I) - J \left(i \frac{\d \bar{\varphi}_I}{\d \tau} \right)
    \right] \nonumber\\
    &= i \int_0^{\Delta \tau_{\rm per}} \d\tau
    \left[
    \frac{m}{2} \left(\frac{\d \bar{r}_I}{\d \tau}\right)^2
    + \frac{m}{2} \bar{r}_I^2 \left(\frac{\d \bar{\varphi}_I}{\d \tau} \right)^2
    + V(\bar{r}_I) - J  \frac{\d \bar{\varphi}_I}{\d \tau} 
    \right] = i S_{R,E} [\bar{r}_I,\bar{\varphi}_I;J]\, .
    \label{eq:S_vertical_raw_new}
\end{align}
Altogether, this now again implies that in the appropriate limits, $-2 {\rm Im}(S_{R,\epsilon}[\bar{r},\bar{\varphi}])=-2 S_{R,E} [\bar{r}_I,\bar{\varphi}_I]=-S_{R,E} [\bar{r}_b,\bar{\varphi}_b]$. Crucially, this allows us to directly evaluate the projection of the steadyon onto $\gamma_\tau$ \textit{without} the need to solve for the generically more complicated form along $\gamma_\epsilon$.

Before generalising these results, we want to highlight an important technical aspect of our discussion: we found that an initial state with a non-vanishing angular momentum inevitably leads to an instanton whose associated cyclical variable, the angle $\varphi$, is strictly imaginary. This is the mechanical realisation of the general fixed-charge prescription previously used in the field-theory literature, e.g., in Refs.~\cite{Lee:1988zz,Coleman:1990as,Lee:1994np,Levkov:2017paj,Barni:2026toappear}. From the point of view of our discussion, this result is indeed not surprising. In the steadyon picture, semi-classical solutions are complex by default. The description in terms of Euclidean-time instantons only arises through a projection, which can indeed lead to an imaginary solution. 

\subsection{Summary and Euclidean-time procedure}
\label{sec:recipe}
Our discussion so far allows us to directly calculate, to leading order, the tunnelling rate out of a state of non-vanishing but fixed energy and angular momentum. We have shown that the resulting expression can be directly evaluated within the familiar Euclidean-time picture, where it takes the form
\begin{align}\label{eq:GammaRecipe}
    \Gamma \sim e^{-2 (S_{R,E}[\bar{r}_I;J]+  E \Delta \tau_{\rm per})} |\psi (r_*,t=0)|^2\, .
\end{align}
If the conserved Noether charge can be easily integrated out, Eq.~\eqref{eq:GammaRecipe} can be evaluated through the following steps. In the following, we denote by $\varphi$ \textit{any} cyclic variable of an arbitrary mechanical system and by $J$ its corresponding Noether charge, while $r\equiv \{ r^1,...,r^N\}$ represents all non-cyclical coordinates.

\begin{enumerate}
    \item Integrate out the cyclic variable by inverting the defining equation of its associated Noether charge, $J\equiv \partial L / \partial \dot{\varphi}$. 
    \item Absorb the Noether charge-dependent term into the potential, $V(r)\to V_{\rm eff}(r)$.
    \item Construct the Euclidean-time Routhian action $S_{R,E}$ as
    \begin{align}\label{eq:SRERecipe}
        S_{R,E}[r;J]=\int_0^{\Delta \tau_{\rm per}} \d \tau \ \frac{m}{2}g_{ij}\dot{r}^i\dot{r}^j+V_{\rm eff}(r)\, ,
    \end{align}
    where the Noether charge is identical to that in the initial state, and $g_{ij}$ denotes the metric in the submanifold spanned by the non-cyclical variables $r$.
    \item Identify the submanifolds $\Sigma_*$ and $\Sigma_s$ through the initial state's energy $E$ using the conditions
    \begin{align}
        V_{\rm eff}(r_*)=E=V_{\rm eff}(r_s), \quad {\rm for\ all}\quad r_* \in \mathcal{F} \quad {\rm and}\quad  r_s \in \mathcal{R}\, .
    \end{align}
    Recall that $\mathcal{F}$ was defined as the basin of attraction of the false vacuum, while $\mathcal{R}$ denotes the region into which the particle tunnels. See Fig.~\ref{fig:sombrero}.
    \item Solve the Euclidean equation of motion for $r$ with vanishing initial velocity,
    \begin{align}
        \frac{\d^2 g_{ij }\bar{r}^j_I}{\d \tau^2}= \frac{\partial V_{\rm eff}(\bar{r}_I)}{\partial r^i}, \qquad \bar{r}_I(0)\in \Sigma_* \qquad  \frac{\d \bar{r}_I}{\d \tau}(0)=0\, .
    \end{align}
    \item Read off $\Delta \tau_{\rm per}$ as the Euclidean time when $\bar{r}_I$ reaches $r_s$ for the \textit{first} time.
    \item Evaluate $\bar{S}_E$ in Eq.~\eqref{eq:SRERecipe} as the Euclidean-time Routhian action $S_{R,E}$ of $\bar{r}_I$.
\end{enumerate}

\subsection{Detailed example} \label{sec:ex}
\label{sec:example}
To make the general discussion fully explicit, we now consider the simplest non-trivial model: a particle of unit mass in two dimensions, subject to the rotationally invariant potential
\begin{equation}
V(r)= \frac{1}{2}m \omega^2r^2-\frac{\lambda}{4}r^4 ,\qquad V_{\rm eff}(r)=V(r)+{J^2\over 2mr^2}\, ,
\label{eq:quartic_potential_example}
\end{equation}
where, for simplicity, in the explicit examples below we choose the parameters such that $\omega/m=\lambda/m^5=1$ so that energies are measured in units of \(m\), while distances and times are measured in units of \(m^{-1}\). In the same conventions, the angular momentum is dimensionless, since $J = m r^2 \dot\alpha$ and therefore \([J]=1\) in natural units, where \(\alpha\) itself is dimensionless.
As in the previous sections, it is convenient to work in polar coordinates $(r,\alpha)$, so that the conserved quantity is the momentum conjugate to the symmetry direction $\alpha$. To allow for a numerical treatment of the steadyons, we furthermore choose the particle's energy and angular momentum in these dimensionless units as $J=0.1$ and $E=0.101m$ (for which $mr_*=0.375$), i.e., only slightly above the value of the effective potential in the false vacuum. This corresponds to well-separated entrance and emergence radii and will ultimately give rise to a numerically well-controlled finite Euclidean time interval.

Following the structure of our discussions so far, we now first construct the relevant steadyon solutions explicitly, keeping track of both variables. We then demonstrate the evaluation of the tunnelling rate directly in Euclidean time.

\begin{figure}[t]
    \centering
    \includegraphics[width=\linewidth]{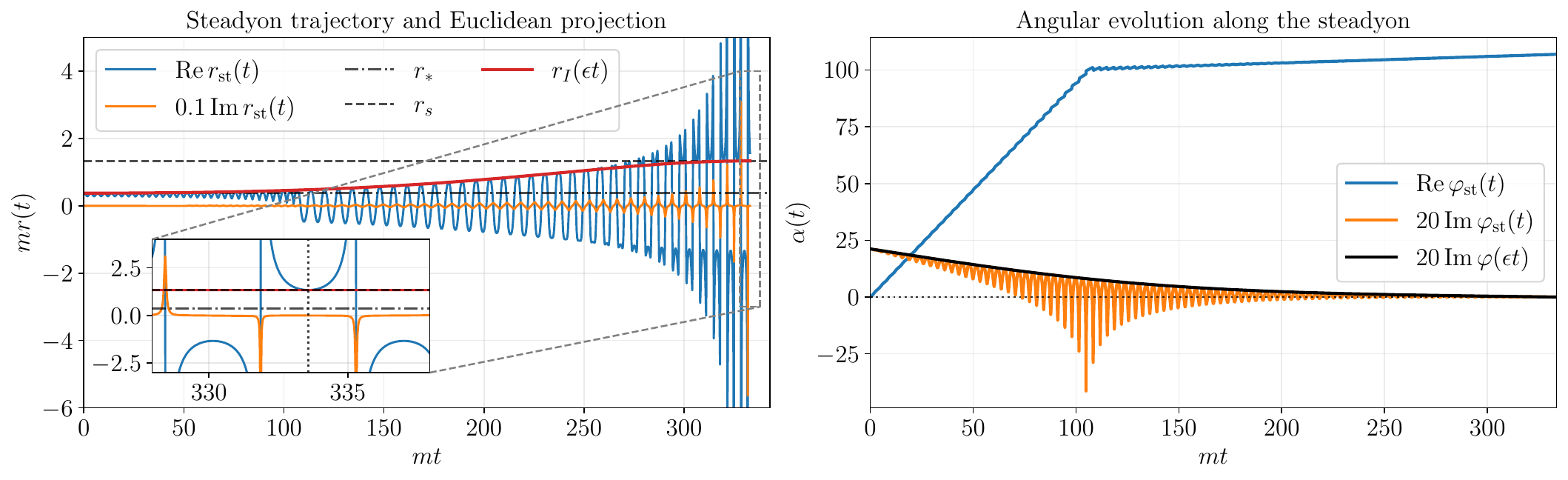}
    \caption{The steadyon solution and the Euclidean instanton for the same fixed angular momentum $J$. \textit{Left}: the radial degree of freedom $\bar{r}(t)$, to be compared with the instanton obtained from Eq.~\eqref{eq:euclidean_1D_concrete_example}, shown in the red solid line. \textit{Right}: the angular degree of freedom $\bar{\varphi}(t)$, together with the Euclidean-time instanton obtained from Eq.~\eqref{eq:alpha_prime_concrete_example}.}
    \label{fig:Steadyons}
\end{figure}

In order to obtain a concrete solution for the steadyon, we first make the additional choice $\epsilon = \frac{1}{100}$. For our concrete potential, the steadyon then emerges as solution to the equations of motion and initial conditions
\begin{gather}
    0=  \ddot{\bar{r}}(t) - \bar{r}(t) \dot{\bar{\varphi}}^2(t) + (1 - 2i \epsilon) (m^2\bar{r}(t)-m^4\bar{r}^3(t)), \quad \bar{r}(0)=r_*, \quad \dot{\bar{r}}(0)=0 \\ 
    \dot{\varphi}(t)=\frac{(1-i\epsilon)\cdot J}{m\bar{r}^2 (t)}, \qquad \bar{\varphi}(\Delta t_{\rm per})=0\, ,
\end{gather}
where the last choice is made possible by the rotational invariance of the system. It is now straightforward to solve these equations numerically, leading to the solutions depicted in Fig.~\ref{fig:Steadyons}. We furthermore determine $\Delta t_{\rm per}$ as the time when the steadyon first satisfies $|\bar{r}(t)-r_s|^2<\epsilon$, yielding $m\Delta t_{\rm per}=333.6$. Using these solutions, we find the exponent of the tunnelling rate
\begin{align}
   \ln (\Gamma)\simeq -2 {\rm Im}(S_{R,\epsilon} [\bar{r},\bar{\varphi};J])+2 E \epsilon \Delta t_{\rm per}= -1.452 + 0.673 = -0.779\, .
\end{align}

\begin{figure}[t]
    \centering
    \includegraphics[width=\linewidth]{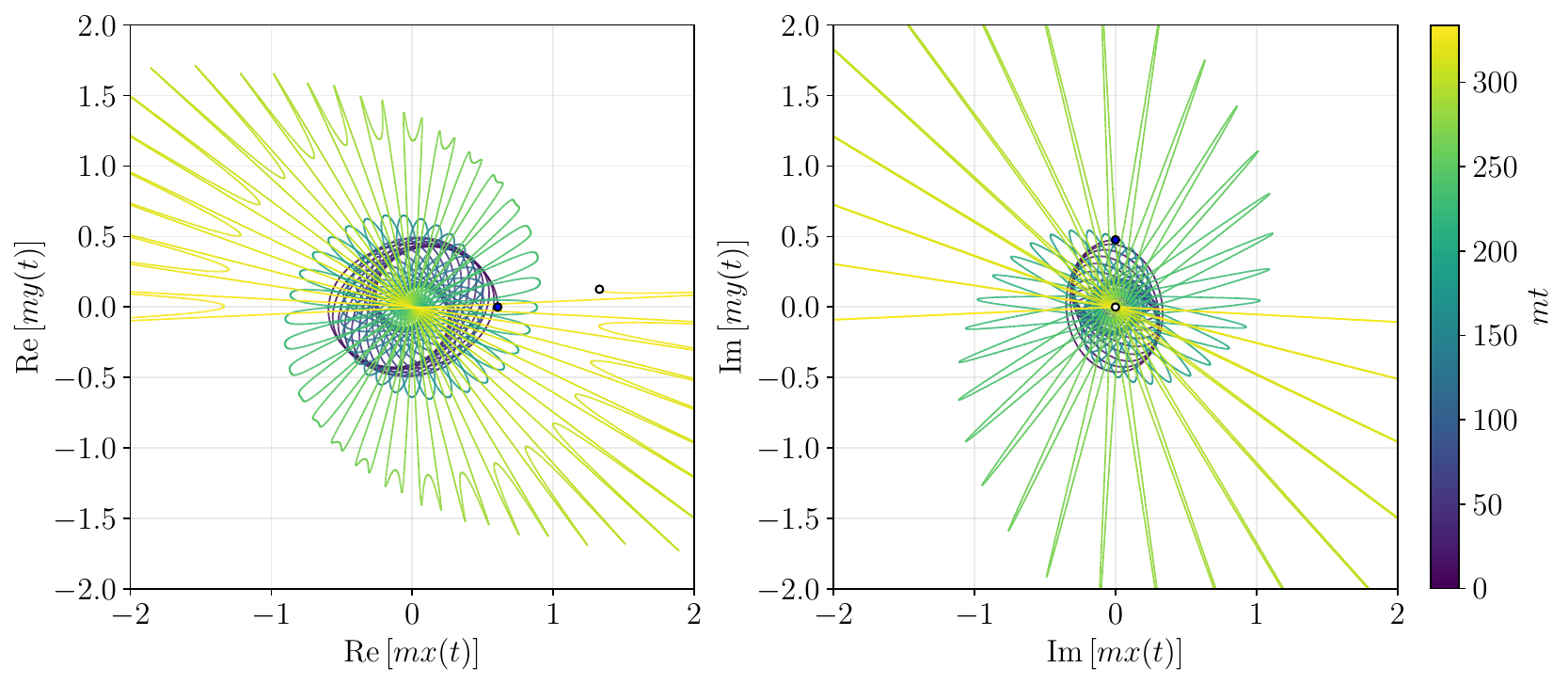}
    \caption{Projection of the steadyon onto the original two-dimensional plane using the Cartesian parametrisation \eqref{eq:cartesian_concrete_example}. \textit{Left}: the real part of the trajectory. \textit{Right}: the imaginary part. These panels are not to be interpreted as literal particle motion in real space; rather, they visualise how a regular saddle in adapted variables becomes a complex configuration when expressed in the original coordinates.}
    \label{fig:Steadyons_trajectories}
\end{figure}

Having a concrete solution available, it is also straightforward to represent it in terms of the Cartesian coordinates
\begin{align}\label{eq:cartesian_concrete_example}
    x=r \, \cos(\varphi)\, , \qquad y=r\, \sin (\varphi)\, .
\end{align}
We present the corresponding trajectories in Fig.~\ref{fig:Steadyons_trajectories}. This representation also highlights an important subtlety of the semi-classical solutions dominating the tunnelling. Neither the steadyon, nor its corresponding instanton describe an actual physical motion. In particular, the final position $(r_s,\varphi_s)$ is \textit{not} to be understood as the position where the particle appears at the time $t$. Instead, these solutions are merely an unphysical technical tool for the evaluation of the path integrals determining the tunnelling rate. The apparently large excursion in Cartesian variables is therefore not a pathology, but simply a reflection of the nature of the semi-classical solutions.

We can now repeat the same calculation directly in Euclidean time, again keeping track of both variables. We can first solve the effective equation of motion for the angular degree of freedom, for which we recover the familiar Euclidean instanton equations in the effective potential,
\begin{align}
    m\frac{\d^2 \bar{r}_I}{\d \tau^2}= V^\prime_{\rm eff}(\bar{r}_I)=m^3\bar{r}_I-m^5\bar{r}_I^3- \frac{J^2}{m\bar{r}_I^3}\, , \qquad \frac{\d \bar{r}_I}{\d \tau}(0)=0\, .\label{eq:euclidean_1D_concrete_example}
\end{align}
The angular variable is then reconstructed from the fixed-$J$ condition,
\begin{equation}
\frac{\d \bar{\varphi}_I}{\d \tau}(\tau)= -i\frac{J}{\bar{r}_I(\tau)^2}\,,
\label{eq:alpha_prime_concrete_example}
\end{equation}
which again highlights the general statement that the variable conjugate to the conserved quantity becomes imaginary in Euclidean time.

\begin{figure}[t]
    \centering
    \includegraphics[width=\linewidth]{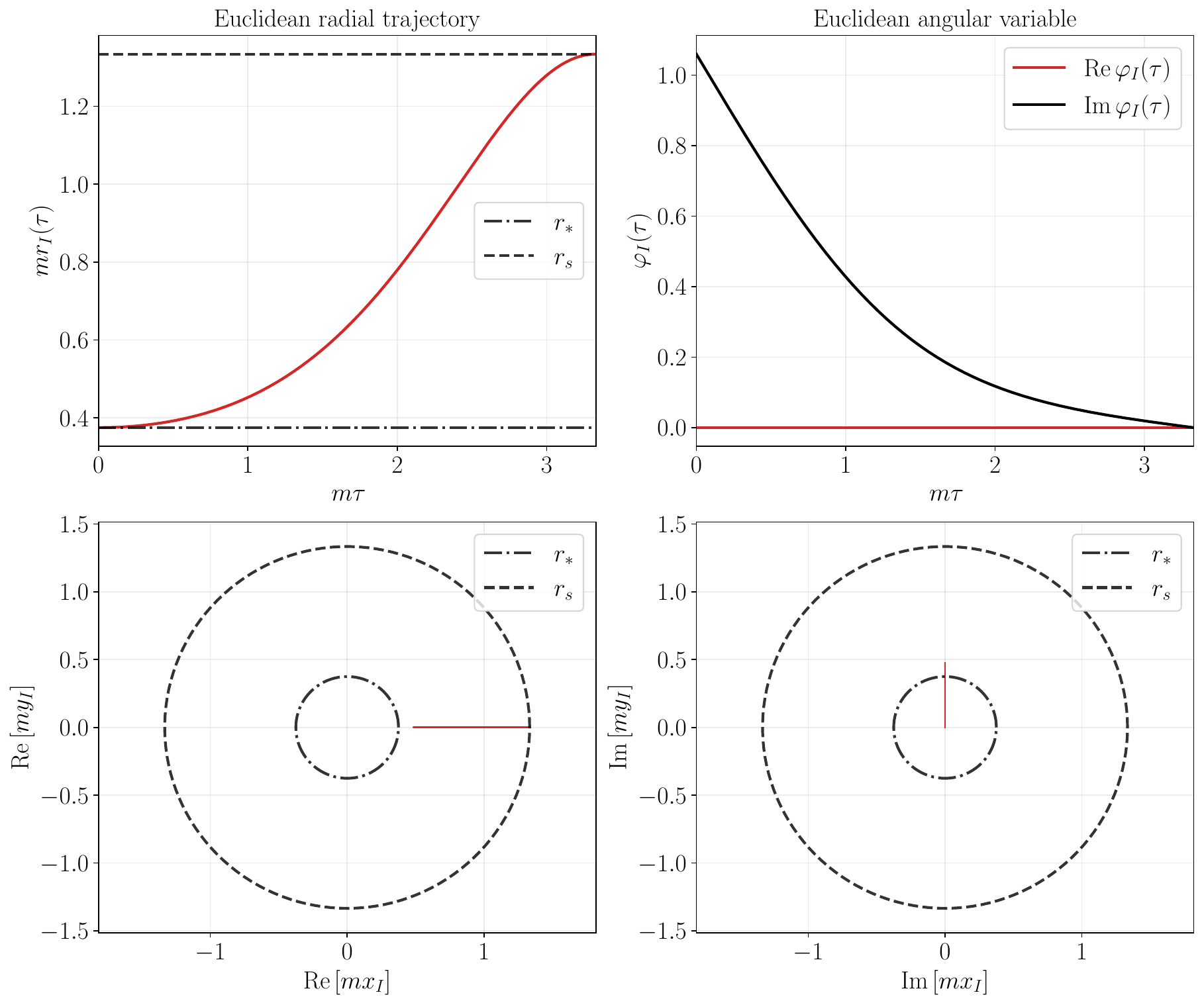}
   \caption{Euclidean fixed-$J$ instanton for the illustrative quartic potential \eqref{eq:quartic_potential_example} at \(J=0.1\). \textit{Upper left}: radial profile \(mr(\tau)\), interpolating from \(r_\ast\) to \(r_s\). \textit{Upper right}: Euclidean angular variable \(\varphi(\tau)\), whose real part vanishes while the imaginary part evolves non-trivially, as implied by the fixed-angular-momentum condition \eqref{eq:alpha_prime_concrete_example}. \textit{Lower panels}: projection of the same solution onto Cartesian variables, showing the real and imaginary parts separately. The dashed circles indicate the radii \(r_\ast\) and \(r_s\).}
    \label{fig:Instantons}
\end{figure}

The resulting Euclidean instanton is shown in Fig.~\ref{fig:Instantons}. The top left panel displays the radial profile $\bar{r}(\tau)$, obtained from the reduced one-dimensional problem~\eqref{eq:euclidean_1D_concrete_example}. The top right panel shows the corresponding angular variable reconstructed from \eqref{eq:alpha_prime_concrete_example}. As anticipated by the general discussion, the radial degree of freedom behaves as an ordinary bounce in the effective potential \eqref{eq:quartic_potential_example}, while the angular variable is purely imaginary. This is the most direct manifestation, in the present model, of the fact that the fixed-$J$ Euclidean saddle does not lie on the naive real slice. As for the steadyon, we determine $m\Delta \tau_{\rm per}=3.329$ as the Euclidean time when the instanton $\bar{r}_I(\tau)$ first reaches $r_s$.

From these Euclidean-time solutions, it is now straightforward to again calculate the exponent of the tunnelling rate. We indeed find agreement with the value obtained through the real-time computation up to a deviation of order $\epsilon$:
\begin{align}
   \ln (\Gamma)\simeq -2 S_{R,E} [\bar{r}_I,\bar{\varphi}_I;J]+2 E \Delta \tau_{\rm per} = -1.451 + 0.672 = -0.779\, .
\end{align}

\section{Generalisation to field theory}
\label{sec:steadyon_field_theory}

In the following, we generalise our earlier arguments to the simple field-theoretic example of a complex scalar field $\phi$ subject to a global $U(1)$ symmetry, described by the action
\begin{equation}
S[\phi,\phi^*] = \int d^4x\, \left[ \partial_\mu\phi^*\,\partial^\mu\phi- V(\phi^*\phi)\right]\, .
\label{eq:S_field_basic_new}
\end{equation}
The symmetry becomes manifest in the local polar parameterisation
\begin{equation}
\phi(x)=\frac{1}{\sqrt2}\rho(x)e^{i\alpha(x)}\, ,
\label{eq:phi_rho_alpha_new}
\end{equation}
in terms of which the action reads
\begin{equation}
S[\rho,\alpha]=\int d^4x\,\left[\frac12\,\partial_\mu\rho\,\partial^\mu\rho + \frac{\rho^2}{2}\,\partial_\mu\alpha\,\partial^\mu\alpha - V\!\left(\frac{\rho^2}{2}\right) \right]\, .
\label{eq:S_rhoalpha_new}
\end{equation}
The associated Noether current is
\begin{equation}
j^\mu=\frac{\partial\mathcal L}{\partial(\partial_\mu\alpha)}=\rho^2\partial^\mu\alpha\, , \qquad \partial_\mu j^\mu=0\, ,
\label{eq:Noether_current_field_new}
\end{equation}
and the conserved charge is
\begin{equation}
Q=\int d^3x\,j^0(x)\, .
\label{eq:Q_field_new}
\end{equation}
In the remainder of this section, we first discuss how our earlier derivation can be generalised to the case of quantum fields in the language of wave functionals. Then, in Sec.~\ref{sec:QFTSC}, we focus on the semi-classical evaluation of the tunnelling rate.

\subsection{Field wave functional}

Starting from an initial state of definite charge $Q$, the relevant amplitude may be written schematically as
\begin{equation}
A_Q[\phi_s, \Delta t|\Sigma_*] = \int_{\Sigma_*} \mathcal D\phi_i\, \Psi_Q[\phi_i]\, D_F[\phi_s,\Delta t |\phi_i]\, ,
\label{eq:AQ_functional_start}
\end{equation}
where $\Psi_Q[\phi_i]$ is the initial wave functional, $D_F$ the usual real-time propagator and $\Sigma_*$ the field-theoretic generalisation of its mechanical counterpart. The exact analogue of the mechanical angular coordinate is not the local phase field $\alpha(\mathbf x)$, but rather the \emph{global} coordinate $\Theta$ along the $U(1)$ orbit in the full space of field configurations. Locally, one may therefore introduce adapted coordinates on configuration space of the form
\begin{equation}
\phi_i(\mathbf x)=e^{i\Theta_i}\,\chi_i(\mathbf x;Y)\, ,
\label{eq:config_space_coords_field}
\end{equation}
where $\Theta_i$ parametrises the global $U(1)$ orbit and $Y^A$ denotes the remaining coordinates transverse to it. This decomposition is conceptually different from \eqref{eq:phi_rho_alpha_new}: the pair $(\rho,\alpha)$ consists of local fields defined pointwise in physical space, whereas $(\Theta,Y^A)$ are coordinates on the infinite-dimensional space of field configurations. Since the conserved charge \(\hat Q\) is the generator of the global \(U(1)\) symmetry, a finite rotation of the field configuration by \(\Theta_i\) acts on the state through \(U(\Theta_i)=e^{i\Theta_i \hat Q}\). For an initial state in a fixed-charge sector, \(\hat Q\ket{\Psi_Q}=Q\ket{\Psi_Q}\), this immediately implies
\begin{equation}
\Psi_Q[\phi_i]=\Psi_Q[e^{i\Theta_i}\chi_i] = e^{iQ\Theta_i}\,\psi_Q[Y_i]\, .
\label{eq:wave functional_fixedQ_field}
\end{equation}
Thus, in complete analogy with the factor $e^{iJ\phi_*}$ in quantum mechanics, the exponent acquires a boundary term proportional to $Q\Theta_i$, and the corresponding fixed-charge functional is the configuration-space Routhian
\begin{equation}
S_R=S-\int dt\,Q\,\dot\Theta\, .
\label{eq:abstract_routhian_field}
\end{equation}
At this abstract level, the analogy with the mechanical problem is exact. The transverse coordinates $Y^A$ are, however, highly non-local functionals of the fields, which can in general not be constructed explicitly. More details can be found in Appendix~\ref{app:config_space_routhian}. 

An essential feature of the semi-classical approach is that it allows for leading-order calculations without detailed knowledge of the wave functional. While in this section, we only focus on the most straightforward case in which the charge is directly encoded in the initial state, it is important to keep in mind that this scenario is not necessarily universal. In finite-temperature systems, for instance, the initial state would be described by a density matrix, while the charge would be enforced through an additional projector involving a Lagrange multiplier~\cite{Coleman:1990as,Levkov:2017paj,Barni:2026toappear}. Towards the end of this section, we briefly discuss the connection between our prescription and such a Lagrange-multiplier based formulation.

\subsection{Tunnelling at fixed $Q$ in QFT}\label{sec:QFTSC}

To highlight the effects of the Noether charge, we restrict ourselves to tunnelling out of a state with minimal energy but non-vanishing charge $Q$. This is the direct analogue of a local minimum in a fixed-$Q$ sector, and allows us to set
\begin{equation}
E=0\, , \qquad \Delta\tau_{\rm per}\to\infty\, .
\label{eq:E_zero_limit_field}
\end{equation}
The choice \(E=0\) ensures that the Euclidean action remains finite in the infinite-period limit.\footnote{For discussions of the direct approach and its generalisation to field theory, see Refs.~\cite{Andreassen:2016cff,Andreassen:2016cvx,Steingasser:2022yqx}.}

A natural example of such an initial state is a metastable Q-ball. Its stationary form is
\begin{equation}
\phi(t,\mathbf x)=\frac{\rho(r)}{\sqrt2}\,e^{i\omega t}\, ,
\label{eq:qball_ansatz_new}
\end{equation}
where \(\omega\) plays the role of the quantity conjugate to the conserved charge. The corresponding fixed-charge energy functional may be written as
\begin{equation}
E_\omega = E+\omega\left[ Q-\int d^3x\,j^0(x) \right]\, .
\label{eq:Eomega_field_new}
\end{equation}
Extremising \(E_\omega\) is equivalent to extremising the usual three-dimensional reduced functional
\begin{equation}
S_3[\rho;\omega] = \int d^3x\, \frac12(\nabla\rho)^2
+ V\!\left(\frac{\rho^2}{2}\right) - \frac{\omega^2}{2}\rho^2\,  ,
\label{eq:S3_qball_new}
\end{equation}
whose equation of motion is
\begin{equation}
\rho''+\frac{2}{r}\rho' - \frac{\partial}{\partial \rho} \left[ V\!\left(\frac{\rho^2}{2}\right) - \frac{\omega^2}{2}\rho^2
\right] =0\, , \qquad \rho'(0)=0\, , \qquad \rho(r\to\infty)\to 0\, .
\label{eq:qball_profile_eom_new}
\end{equation}
We denote the corresponding solution by \(\rho_*\). For our purposes, this configuration plays the same role as \(r_*\) in the mechanical problem. The target set \(\Sigma_s\) is now replaced by a submanifold of field configurations whose reduced energy functional vanishes,
\begin{equation}
0=U[\varphi_s]=\int_{\mathbb{R}^3}\d x\ |\nabla\varphi_s|^2+V(|\varphi_s|^2)\, , \qquad \forall\,\varphi_s\in\Sigma_s\, .
\label{eq:Sigma_s_field_new}
\end{equation}
The initial charge density is then
\begin{equation}
j^\mu\dot{=}(\omega\,\rho_*^2(r),0,0,0)\, ,
\label{eq:initial_charge_density_field_new}
\end{equation}
exactly mirroring the role of the fixed angular momentum density in the mechanical example.

For practical calculations, it is convenient to replace the abstract configuration-space Routhian by an equivalent local formulation in terms of the fields \(\rho\) and \(\alpha\). This is achieved by introducing the quantity conjugate to the fixed charge as a Lagrange multiplier directly in the path integral. In other words, the local field-theory version of the fixed-\(Q\) Routhian is nothing but the constrained action obtained by enforcing the charge condition
\begin{equation}
Q=\int d^3x\,\rho^2\,\dot\alpha\, ,
\label{eq:charge_constraint_field_local}
\end{equation}
with a Lagrange multiplier $\omega$. In the present context, $\omega$ plays the same role as in the mechanical discussion: it is the variable conjugate to the conserved charge, and for the stationary Q-ball background it coincides with the frequency appearing in \eqref{eq:qball_ansatz_new}.

Starting from the regulated real-time action
\begin{equation}
S_\epsilon[\rho,\alpha] = \int dt\,d^3x\,  \frac12\,\dot\rho^{\,2} +\frac{\rho^2}{2}\,\dot\alpha^{\,2}-(1-2i\epsilon)
\left( \frac12\,(\nabla\rho)^2 +\frac{\rho^2}{2}\,(\nabla\alpha)^2
+ V\!\left(\frac{\rho^2}{2}\right) \right)\, ,
\label{eq:Seps_field_local_revised}
\end{equation}
a convenient local implementation of the fixed-charge condition is obtained by introducing the quantity conjugate to $Q$ as a Lagrange multiplier directly in the path integral. In this representation, one works with the constrained functional
\begin{equation}
S_{R,\epsilon}[\rho,\alpha;\omega] = S_\epsilon[\rho,\alpha]
+ \int dt\,\omega \left[ Q-\int d^3x\,\rho^2\,\dot\alpha \right]\, .
\label{eq:SR_eps_field_local_revised}
\end{equation}
Thus, the local field-theory version of the fixed-\(Q\) Routhian is nothing but the regulated action supplemented by the Lagrange multiplier term imposing the charge constraint. Explicitly, after completing the square in the time derivative of the phase,
\begin{align}
S_{R,\epsilon}[\rho,\alpha;\omega] = \int d^4x\, \Bigg[ \frac12\,\dot\rho^{\,2} +\frac{\rho^2}{2}\,(\dot\alpha-\omega)^2
-\frac{\rho^2}{2}\,\omega^2 -(1-2i\epsilon) \left( \frac12\,(\nabla\rho)^2 +\frac{\rho^2}{2}\,(\nabla\alpha)^2 + V\!\left(\frac{\rho^2}{2}\right) \right) \Bigg] +\int dt\,\omega\,Q\, .
\label{eq:SR_eps_field_local_square_revised}
\end{align}
Varying with respect to $\omega$ reproduces the fixed-charge condition. Varying with respect to $\rho$ and $\alpha$ then yields the local steadyon equations
\begin{gather}
0= \ddot{\bar\rho}(x) - \bar\rho(x)\,\big(\dot{\bar\alpha}(x)-\omega\big)^2 + \bar\rho(x)\,\omega^2 + (1-2i\epsilon)
\left[ -\Delta\bar\rho(x) + \bar\rho(x)\,(\nabla\bar\alpha)^2
+ \bar\rho(x)\, V'\!\left(\frac{\bar\rho^2(x)}{2}\right)
\right]\, ,
\label{eq:SPrhobar_new_revised}
\\[0.5em]
0= \partial_t\!\left[\bar\rho^2(x)\,\big(\dot{\bar\alpha}(x)-\omega\big)\right] - (1-2i\epsilon)\, \nabla\!\cdot\!\left[\bar\rho^2(x)\,\nabla\bar\alpha(x)\right]\, ,
\label{eq:SPalphabar_new_revised}
\\[0.5em]
\omega\,\rho_*^2(r) = (1+i\epsilon)\, \left[\dot{\bar\alpha}(x)\,\bar\rho^2(x)\right]_{t=0}\, ,
\label{eq:SPalpha_bc_new_revised}
\end{gather}
together with
\begin{equation}
\bar\rho(x)\big|_{t=0}=\rho_*(x)\, , \qquad \dot{\bar\rho}(x)\big|_{t=0}=0\, , \qquad \bar\rho(x)\big|_{t\to\infty}\in\Sigma_s\, .
\label{eq:SP_bc_field_new_revised}
\end{equation}
To leading order, the tunnelling rate is controlled by the imaginary part of the corresponding complex Routhian action,
\begin{equation}
\Gamma \sim \exp\!\left[-2\,{\rm Im}\,S_{R,\epsilon}[\bar\rho,\bar\alpha;\omega]\right]\, .
\label{eq:Gamma_SR_eps_field_new_revised}
\end{equation}
Exactly as in the mechanical problem, the same quantity may be evaluated more directly in Euclidean time by projecting the steadyon onto the contour \(\gamma_\tau\), see Sec.~\ref{sec:euclidean_picture}. In the fixed-charge representation, the corresponding Euclidean Routhian action is
\begin{equation}
S_{E,R}[\rho,\beta;\omega] = S_E[\rho,\beta] + \int d\tau\,\omega
\left[ Q-\int d^3x\,\rho^2\,\partial_\tau\beta \right]\, ,
\label{eq:SER_local_field_revised}
\end{equation}
with
\begin{equation}
S_E[\rho,\beta] = \int d\tau\,d^3x\, \frac12(\partial_\tau\rho)^2 +\frac12(\nabla\rho)^2 +\frac{\rho^2}{2}(\partial_\tau\beta)^2 +\frac{\rho^2}{2}(\nabla\beta)^2 + V\!\left(\frac{\rho^2}{2}\right)\, .
\label{eq:SE_rhobeta_new_revised}
\end{equation}
After completing the square,
\begin{equation}
    V\!\left(\frac{\rho^2}{2}\right) \to V\!\left(\frac{\rho^2}{2}\right) - {\omega^2 \over 2}\rho^2\, .
\end{equation}
Varying with respect to $\omega$ gives
\begin{equation}
Q=\int d^3x\,\rho^2\,\partial_\tau\beta\, .
\label{eq:charge_constraint_E_field_revised}
\end{equation}
The Euclidean equations are then
\begin{align}
0&= -\partial_\tau^2\rho -\nabla^2\rho +\rho\big(\partial_\tau\beta-\omega\big)^2 -\rho\,\omega^2
+\rho(\nabla\beta)^2 + \rho\, V'\!\left(\frac{\rho^2}{2}\right)\, ,
\label{eq:eom_rho_beta_E_new_revised}
\\
0&= \partial_\tau\!\left[\rho^2\big(\partial_\tau\beta-\omega\big)\right] + \nabla\!\cdot\!\left(\rho^2\,\nabla\beta\right)\, ,
\label{eq:eom_beta_E_new_revised}
\end{align}
with boundary conditions
\begin{equation}
\rho(r\to\infty)\to 0\, , \qquad \rho'(0)=0\, , \qquad \partial_\tau\beta(0)=\omega\, , \qquad \beta(0)=\beta_*\, .
\label{eq:beta_bc_new_revised}
\end{equation}
In the notation of the original complex phase, this corresponds to
\begin{equation}
\bar\alpha_I(\tau,\mathbf x)=i\,\bar\beta_I(\tau,\mathbf x)\, ,
\qquad \partial_\tau\bar\alpha_I(0)=-\,i\omega\, .
\label{eq:alpha_beta_definition_new_revised}
\end{equation}
The Euclidean twist is
\begin{equation}
\eta = \Delta\beta_I = \beta_I(+\infty)-\beta_I(-\infty)\, ,
\label{eq:eta_field_new_revised}
\end{equation}
so that the Euclidean Routhian may also be written as
\begin{equation}
S_{E,R}[\rho_I,\beta_I] = S_E[\rho_I,\beta_I]+\eta Q\, .
\label{eq:SER_field_new_revised}
\end{equation}
The leading semiclassical exponent, therefore, takes the form
\begin{equation}
\Gamma \sim \exp\!\left[-2\,S_{E,R}[\rho_I,\beta_I]\right] = \exp\!\left[-2\left(S_E[\rho_I,\beta_I]+\eta Q\right)\right]\, .
\label{eq:Gamma_Euclidean_field_new_revised}
\end{equation}
To connect with the notation commonly used in the fixed-charge literature, we introduce the operator $\ostar$ as
\begin{equation}
\bar\phi_I(x)=\frac{\bar\rho_I(x)}{\sqrt2}e^{\bar\beta_I(x)}\, ,
\qquad \bar\phi_I^\ostar(x)=\frac{\bar\rho_I(x)}{\sqrt2}e^{-\bar\beta_I(x)}\, .
\label{eq:phi_phiostar_new_revised}
\end{equation}
In terms of these fields, the Euclidean action takes the compact form
\begin{equation}
S_E = \int d\tau\,d^3x\, \left[ \partial_\mu \bar\phi_I\,\partial_\mu \bar\phi_I^\ostar + V(\bar\phi_I^\ostar\bar\phi_I)
\right]\, .
\label{eq:SE_phi_phiostar_new_revised}
\end{equation}
Accounting for differences in notation, this reproduces the same complex Euclidean ansatz encountered in earlier fixed-charge analyses such as Refs.~\cite{Lee:1994np,Levkov:2017paj, Barni:2026toappear}.

Although we have used a metastable Q-ball as our reference example, the same construction applies much more generally. In particular, one may equally well start from a homogeneous charged background, or from any other metastable configuration carrying a fixed global charge, and study the nucleation of a charged bubble under the same constraint. See, for example, \cite{Lee:1994np,Barni:2026toappear}. Crucially, all these examples are captured by our formalism, differing exclusively through the form of the initial profile $\rho_*$ and the target set $\Sigma_s$. 

\section{Conclusions}
\label{sec:conclusion}

We have studied tunnelling out of initial states carrying a conserved Noether charge in both quantum mechanics and quantum field theory. By relying on the steadyon picture based on the direct approach, we were able to derive the tunnelling rate entirely from first principles and evaluate the resulting real-time path integrals. Our derivation makes fully transparent the regimes of validity of our result as well as the underlying assumptions. Moreover, it offers a clear understanding of the emergence of complex saddle points found by earlier works. We find, in particular, that this is a generic feature of tunnelling out of charged states.

We first performed the complete calculation in full detail for the simplest suitable mechanical system, a point particle in two spatial dimensions carrying a conserved angular momentum. On a formal level, this amounts to a rotationally invariant potential and an initial wave function which depends on the angular variable exclusively through a phase. When integrating over the suitable starting points of semi-classical tunnelling process, this phase manifests through an initial condition on the steadyon's angular velocity, allowing us to relate its angular momentum to the corresponding quantum number of the initial state. When performing the integral over the suitable end points of the semi-classical dynamics, the same factor can be recast as a correction to the potential, allowing us to recover the familiar angular momentum barrier and corresponding effective potential. Crucially, by evaluating the relevant path integrals through this real-time technique, we were able to simultaneously account for the initial angular momentum and allow for energies elevated above the false vacuum of the effective potential. Following earlier works on the steadyon picture, we have shown that our result can be rephrased in familiar Euclidean-time language. We found that on this level, the tunnelling process can be described entirely in terms of the radial motion upon replacement of the Euclidean action by its Routhian counterpart, which arises naturally from our earlier discussions. This leads to a simple prescription for the calculation of tunnelling rates which can be easily applied to many systems.

In field theory, the conserved quantity is not a local mechanical momentum, but a spatially integrated Noether charge of the form
\begin{equation}
    C=\int d^{d-1}x\, j^0(x)\, .
\end{equation}
Using the language of wave functionals, we have shown that many of our calculations can nevertheless be applied in a straightforward manner in this new context. The main difference occurs only during the semi-classical evaluation of the relevant path integrals, where the non-local form of the conserved charge prevents a direct elimination of the cyclical degree of freedom in favour of this charge. We have shown that this problem can be circumvented through the use of an integral constraint. Finally, just as for the quantum mechanical case, we have shown that the tunnelling rate can equivalently be understood in the familiar Euclidean-time picture.

Our results point towards several natural applications of significant phenomenological relevance. Our discussion in Sec.~\ref{sec:steadyon_field_theory}, for instance, can be immediately applied to the analysis of vacuum decay in the presence of metastable charge solitons such as Q-balls \cite{Levkov:2017paj}. It moreover offers a path towards the detailed understanding of phase transitions in dense relativistic matter, where conserved charges can be expected to have a significant impact \cite{Barni:2026toappear}. In systems such as neutron stars, proto-neutron stars, or merger remnants, one expects nucleation and phase conversion to occur in media characterised by conserved baryon number, lepton number, or other approximately conserved composition variables on the relevant time scales. In such situations, the appropriate semiclassical problem is precisely of the type studied here, rather than the standard neutral false-vacuum bounce.

\vskip 10pt

{\it Acknowledgements.}
We thank Jos\'e-Ramon Espinosa and Bruno Scheihing-Hitschfeld for helpful discussions.

GB is supported by the grant CNS2023-145069 funded by MICIU/AEI/10.13039/ 501100011033 and by the European Union NextGenerationEU/PRTR. He also acknowledges support from the Spanish Agencia Estatal de Investigaci\'on through the grant ``IFT Centro de Excelencia Severo Ochoa CEX2020-001007-S''.

TS acknowledges partial financial support from the Spanish Agencia Estatal de Investigaci\'on through the grants ``IFT Centro de Excelencia Severo Ochoa CEX2020-001007-S'' and PID2022-137127NB-I00, funded by MCIN/AEI/10.13039/501100011033/FEDER, UE. This project has also received support from the European Union's Horizon 2020 research and innovation programme through the Marie Sklodowska-Curie Staff Exchange grant agreement No.~101086085--ASYMMETRY.

\appendix

\section{WKB approximation for the wave function}
\label{app:wkb}

Our derivation of the tunnelling rate in Sec.~\ref{sec:FPderivation} relied on the condition that the wave function of the considered particle is localised within the false vacuum basin $\mathcal{F}$ through $r<r_*$. In this appendix, we use the WKB approximation to show that for a resonance state with energy $E$ and angular momentum $J\cdot \hbar$, this point can be defined through the condition $V_{\rm eff}(r_*)=E$. To make fully transparent the importance of the semi-classical limit, we will restore $\hbar$ throughout this discussion.

As established in Sec.~\ref{sec:FPderivation}, the wave function of the state $|E,J\rangle$ is of the form
\begin{align}
    \Psi_{E,J} (r, \varphi)= e^{i J \varphi } \psi_{E,J} (r)\, .
\end{align}
The radial part of the wave function, $\psi_{E,J} (r)$, is an approximate eigenfunction of the dimensionally reduced Hamiltonian
\begin{align}
    \hat{H}_J= - \frac{\hbar^2}{2m} \Delta_r+ \frac{\hbar^2 J^2}{2m r^2} + V(r) \equiv  - \frac{\hbar^2}{2m} \Delta_r +V_{\rm eff}(r)\, ,
\end{align}
where the radial component of the Laplace operator is given by
\begin{align}
    \Delta_r = \partial_r^2+ \frac{1}{r}\partial_r\, ,
\end{align}
so that the (approximate) Schr\"odinger equation for the wave function $\psi_{E,J} (r)$ becomes
\begin{align}\label{eq:Schrodinger1D}
     - \frac{\hbar^2}{2m} \left(\psi_{E,J}^{\prime \prime} + \frac{1}{r} \psi_{E,J}^\prime \right) +V_{\rm eff}(r) \cdot \psi_{E,J}\simeq E \cdot \psi_{E,J}\, .
\end{align}
This equation can be further simplified by introducing a rescaled wave function $\rho_{E,J}(r)\equiv \sqrt{r} \cdot \psi_{E,J}(r)$. In terms of this new function, Eq.~\eqref{eq:Schrodinger1D} can be rewritten as
\begin{align}\label{eq:Schrodinger1Drho}
     - \frac{\hbar^2}{2m} \rho_{E,J}^{\prime \prime} + \left( V(r) + \frac{\hbar^2(J^2-1/4)}{2m r^2} \right)  \cdot \rho_{E,J}  =  E \cdot \rho_{E,J}\, .
\end{align}
This equation can now be identified with the Schr\"odinger equation for a resonance state in one spatial dimension, subject to the modified potential
\begin{align}
    \tilde{V}(r)=V(r) + \frac{\hbar^2(J^2-1/4)}{2m r^2}= V_{\rm eff} (r) - \frac{\hbar^2}{8 m r^2}\, .
\end{align}
In the semi-classical limit of interest for our purpose, we can now neglect the last term, leading us to recover the one-dimensional WKB picture for the rescaled wave function $\rho_{E,J}$. Eq.~\eqref{eq:Schrodinger1Drho} can therefore be solved through the usual technique, which we review for completeness.

We start by choosing the Ansatz
\begin{align}\label{eq:WKBansatz}
    \rho_{E,J}(r)= \mathcal{N}_{E,J} e^{i \alpha_{E,J}(r)/\hbar}\, ,
\end{align}
with some normalisation constant $\mathcal{N}_{E,J}$. Through this Ansatz, Eq.~\eqref{eq:Schrodinger1Drho} can be rephrased as an equation for the phase function $\alpha(r)$,
\begin{align}
    i \hbar \alpha_{E,J}^{\prime \prime} - \left( \alpha_{E,J}^\prime \right)^2  = 2m (V_{\rm eff}(r)-E)\, .\label{eq:WKBalpha}
\end{align}
The semi-classical limit then suggests to expand $\alpha_{E,J}$ as a series in $\hbar$, $\alpha_{E,J} = \alpha_{E,J}^0 + \hbar \alpha_{E,J}^1 +...$. Doing so allows us to translate Eq.~\eqref{eq:WKBalpha} to a straightforward-to-solve differential equation for $\alpha_0$,
\begin{gather}
    (\alpha_{E,J}^0)^\prime = \sqrt{2m \left( E- V_{\rm eff}(r) \right)}\\
    \Rightarrow \alpha_{E,J}^0(r)= \pm \int \d r \ \sqrt{2m \left( E-V_{\rm eff}(r) \right)}= \pm i \int \d r \ \sqrt{2m \left( V_{\rm eff}(r)-E \right)}\, .
\end{gather}
Together with Eq.~\eqref{eq:WKBansatz}, this is sufficient to determine the leading-order behaviour of the wave function,
\begin{align}
    \psi_{E,J}(r)\sim \begin{cases}
        \displaystyle
        C_1 \exp\left[\frac{i}{\hbar}\int\d r  \sqrt{2m \left( E-V_{\rm eff}(r) \right)}\right]
            + C_2 \exp\left[-\frac{i}{\hbar}\int\d r  \sqrt{2m \left( E-V_{\rm eff}(r) \right)}\right] & 0<r < r_*
        \\[4mm]
        \displaystyle
        C_3 \exp\left[-\frac{1}{\hbar}\int\d r \sqrt{2m \left(V_{\rm eff}(r) -E \right)}\right] &  r_*<r<r_s
        \\[4mm]
        \displaystyle
        C_4 \exp\left[\frac{i}{\hbar}\int\d r  \sqrt{2m \left( E-V_{\rm eff}(r) \right)}\right] & r_s<r
    \end{cases}\, ,\label{eq:WKBres}
\end{align}
where we have neglected next-to-leading terms as well as corrections to this form near the turning points $r_*$ and $r_s$. Eq.~\eqref{eq:WKBres} is indeed sufficient to understand the localisation of the wave function: for $r<r_*$, the leading-order radius dependence of the wave function is restricted to its phase, allowing for a large probability within this region. For $r_*<r<r_s$, the wave function decays exponentially, with the semi-classical limit corresponding to a parametrically large exponent. Thus, the wave function can, again in the semi-classical limit, be considered to be localised within the region $r<r_*$.

\section{Fixed-charge tunnelling in configuration-space coordinates}
\label{app:config_space_routhian}

In this appendix, we reformulate the fixed-charge tunnelling problem on a more abstract and geometric level, making fully explicit its analogy with the quantum-mechanical discussion. The basic idea is to view a relativistic field theory as a mechanical system on an infinite-dimensional configuration space. From this perspective, the treatment of a fixed conserved charge follows the same structural logic as the treatment of a fixed angular momentum in ordinary mechanics. In particular, one may locally introduce coordinates on configuration space adapted to the symmetry orbits, separating the direction generated by the symmetry from the transverse directions. This perspective is standard in geometric formulations of field theory and is closely related, in a more physics-oriented language, to the use of collective coordinates and moduli-space variables. For general discussions along these lines, see Refs.~\cite{Storchak:2018qbv,Tong:2005un}.

Throughout this appendix, we assume that the initial state is already in a sector of fixed conserved charge $Q$. Thus, unlike the projector-based or constrained-functional derivations used in the main text, the fixed-$Q$ condition is now encoded directly in the wave functional of the initial state. We will show in the following that this naturally leads to the emergence of the analogue of the Routhian, in complete parallel with the quantum-mechanical case.

\subsection{Amplitude in terms of the initial wave functional}

We consider a complex scalar field with action
\begin{equation}
S[\phi,\phi^*]=\int d^4x\, \partial_\mu \phi^*\partial^\mu \phi- V(\phi^*\phi)\, .
\label{eq:appA_action_phi}
\end{equation}
The transition amplitude from an initial state of fixed charge $Q$ to a final field configuration $\phi_s$ at time $\Delta t$ can be written as
\begin{equation}
A_Q[\phi_s, \Delta t|\Sigma_*] = \int_{\Sigma_*} \mathcal D\phi_i\, \Psi_Q[\phi_i]\, D_F[\phi_s,\Delta t |\phi_i]\, ,
\label{eq:appA_AQ_start}
\end{equation}
where $\Psi_Q[\phi_i]$ is the initial wave functional and
\begin{equation}
D_F[\phi_s,\Delta t |\phi_i]= \int_{\phi(0,\mathbf x)=\phi_i(\mathbf x)}^{\phi(\delta t,\mathbf x)=\phi_s(\mathbf x)} \mathcal D\phi\, \exp\!\left(iS[\phi]\right)
\label{eq:appA_kernel}
\end{equation}
is the usual real-time kernel.

The fixed-charge assumption means that $\Psi_Q$ transforms as an eigenfunctional of the global $U(1)$ generator. If the initial configuration is rotated globally as
\begin{equation}
\phi_i(\mathbf x)\to e^{i\theta}\phi_i(\mathbf x)\, ,
\label{eq:appA_global_rotation}
\end{equation}
the wave functional transforms as
\begin{equation}
\Psi_Q[e^{i\theta}\phi_i] = e^{iQ\theta}\, \Psi_Q[\phi_i]\, .
\label{eq:appA_fixedQ_wf}
\end{equation}
This is the field-theory analogue of the factor $e^{iJ\varphi_*}$ in the mechanical problem.

The global $U(1)$ symmetry acts on the full field configuration, not pointwise on an independent mechanical coordinate. Therefore, the natural coordinate conjugate to the total charge $Q$ is not the local phase field $\alpha(\mathbf{x})$, but a global coordinate on configuration space.

To make this explicit, we regard the field theory as a mechanical system on the infinite-dimensional configuration space $\mathcal C$ of field profiles $\phi(\mathbf{x})$. The $U(1)$ symmetry generates one-dimensional orbits in $\mathcal{C}$,
\begin{equation}
\phi(\mathbf x)\mapsto e^{i\Theta}\phi(\mathbf x)\, .
\label{eq:appA_orbit}
\end{equation}
Locally on $\mathcal{C}$, we may introduce coordinates adapted to these orbits:
\begin{equation}
\phi(\mathbf x)=e^{i\Theta}\,\chi(\mathbf x;Y)\, ,
\label{eq:appA_chi_decomp}
\end{equation}
where $\Theta$ parametrizes the global $U(1)$ orbit and $Y^A$ denotes a complete set of coordinates transverse to it. The label $A$ is generally infinite-dimensional.

It is important to distinguish this decomposition from the local polar parametrisation
\begin{equation}
\phi(\mathbf x)=\frac{\rho(\mathbf x)}{\sqrt2}\,e^{i\alpha(\mathbf x)}\, .
\label{eq:appA_local_polar}
\end{equation}
The pair $(\rho,\alpha)$ consists of local fields defined pointwise in physical space. By contrast, $(\Theta,Y^A)$ are coordinates on the space of full field configurations. In particular, $\Theta$ is a single global coordinate, whereas $\alpha(\mathbf{x})$ is a local field. 

Using \eqref{eq:appA_chi_decomp}, the measure over initial configurations can be written locally as
\begin{equation}
\mathcal D\phi_i = d\Theta_i\, \mathcal DY_i\, J[Y_i]\, ,
\label{eq:appA_measure}
\end{equation}
where $J[Y_i]$ is the Jacobian associated with the change of coordinates on configuration space. Eq.~\eqref{eq:appA_fixedQ_wf} then implies that
\begin{equation}
\Psi_Q[\Theta_i,Y_i] = e^{iQ\Theta_i}\,\psi_Q[Y_i]\, .
\label{eq:appA_wf_thetaY}
\end{equation}
The amplitude \eqref{eq:appA_AQ_start} thus becomes
\begin{equation}
A_Q[\phi_s,\Delta t|\Sigma_*] = \int_{\Sigma_i} d\Theta_i\, \mathcal DY_i\, J[Y_i]\, e^{iQ\Theta_i}\,
\psi_Q[Y_i]\, D_F[\phi_s,\Delta t|\Theta_i,Y_i]\, .
\label{eq:appA_AQ_thetaY}
\end{equation}

\subsection{Semiclassical exponent and boundary contribution}

In the semiclassical approximation,
\begin{equation}
D_F[\phi_s,\Delta t|\Theta_i,Y_i] \sim \exp\!\left(iS[\Theta,Y]\right)\, ,
\label{eq:appA_kernel_saddle}
\end{equation}
where $S[\Theta,Y]$ denotes the action evaluated on the relevant saddle connecting the initial configuration labelled by $(\Theta_i,Y_i)$ to the chosen final configuration. The amplitude then takes the form
\begin{equation}
A_Q[\phi_s,\Delta t|\Sigma_*] \sim \int d\Theta_i\, \mathcal DY_i\, \exp\!\left( iS[\Theta,Y] +iQ\Theta_i +\log \psi_Q[Y_i] +\log J[Y_i] \right)\, .
\label{eq:appA_AQ_saddle_form}
\end{equation}
The variation with respect to $\Theta_i$ yields
\begin{equation}
0 = \frac{\partial}{\partial \Theta_i} \left( S[\Theta,Y]+Q\Theta_i
\right) = \frac{\partial S}{\partial \Theta_i}+Q\, .
\label{eq:appA_theta_variation}
\end{equation}
Since $\Theta$ is the coordinate along the global symmetry orbit, its canonically conjugate momentum is the total Noether charge. Thus,
\begin{equation}
\frac{\partial S}{\partial \Theta_i} = -\,Q_{\rm sadd}\, ,
\label{eq:appA_boundary_charge}
\end{equation}
and the saddle condition enforces
\begin{equation}
Q_{\rm sadd}=Q\, .
\label{eq:appA_Qmatch}
\end{equation}
Exactly as in the mechanical problem, the term $Q\Theta_i$ is a boundary contribution associated with the cyclic coordinate. Up to a final-state phase that cancels in the tunnelling rate, it may be rewritten as
\begin{equation}
Q\Theta_i = - \int_0^{\Delta t} dt\, Q\, \dot\Theta\, ,
\label{eq:appA_QTheta_to_bulk}
\end{equation}
so that the relevant semiclassical exponent is governed by
\begin{equation}
S_R[\Theta,Y;Q] = S[\Theta,Y]-\int_0^{\Delta t} dt\, Q\,\dot\Theta\, .
\label{eq:appA_SR_def}
\end{equation}
This is the exact analogue of the mechanical Routhian action.

\subsection{Configuration-space Lagrangian and reduced Routhian}

We now derive the explicit configuration-space form of the action. Using
\begin{equation}
\phi(\mathbf x)=e^{i\Theta}\chi(\mathbf x;Y)\, , \qquad \dot\phi = e^{i\Theta}\big(i\dot\Theta\,\chi+\dot\chi\big)\, ,
\label{eq:appA_phi_dot}
\end{equation}
we find
\begin{equation}
|\dot\phi|^2 = \dot\Theta^2\,|\chi|^2 + i\dot\Theta\big(\chi^*\dot\chi-\dot\chi^*\chi\big) + |\dot\chi|^2\, .
\label{eq:appA_absphidot2}
\end{equation}
The spatial gradient does not depend on \(\Theta\),
\begin{equation}
|\nabla\phi|^2=|\nabla\chi|^2\, ,
\label{eq:appA_spatial_grad}
\end{equation}
and similarly $V(\phi^*\phi)=V(\chi^*\chi)$.

The full Lagrangian can therefore be written as
\begin{equation}
L(\Theta,Y;\dot\Theta,\dot Y) = \frac12\,\mathcal I[Y]\,\dot\Theta^2 + \mathcal A_A[Y]\,\dot Y^A\,\dot\Theta + \frac12\,G_{AB}[Y]\,\dot Y^A\dot Y^B - U[Y]\, ,
\label{eq:appA_general_L}
\end{equation}
where
\begin{align}
\mathcal I[Y] &= 2\int d^3x\, \chi^*\chi\, ,
\label{eq:appA_I_def}
\\[0.3em]
\mathcal A_A[Y]\dot Y^A &= i\int d^3x\,\big(\chi^*\dot\chi-\dot\chi^*\chi\big)\, ,
\label{eq:appA_A_def}
\\[0.3em]
\frac12\,G_{AB}[Y]\dot Y^A\dot Y^B &= \int d^3x\, \dot\chi^*\dot\chi\, , 
\label{eq:appA_G_def}
\\[0.3em]
U[Y] &= \int d^3x\,\Big( |\nabla\chi|^2+V(\chi^*\chi) \Big)\, .
\label{eq:appA_U_def}
\end{align}
Here, $U[Y]$ is the static energy functional, while $\mathcal I[Y]$, $\mathcal A_A[Y]$, and $G_{AB}[Y]$ are the natural geometric data on the reduced configuration space.

Since $\Theta$ is cyclic, the conjugate momentum is conserved,
\begin{equation}
\frac{\partial L}{\partial \Theta}=0\, , \qquad \Rightarrow \qquad P_\Theta = \frac{\partial L}{\partial \dot\Theta} = \mathcal I[Y]\dot\Theta+\mathcal A_A[Y]\dot Y^A\, .
\label{eq:appA_Ptheta}
\end{equation}
This momentum is precisely the total Noether charge,
\begin{equation}
P_\Theta=Q\, .
\label{eq:appA_Ptheta_Q}
\end{equation}
At fixed charge, the semi-classical exponent contains the boundary contribution $Q\Theta_i$ inherited from the transformation of the initial wave functional. Up to an analogous final-state term, which is irrelevant for the present discussion, this can be rewritten as
\begin{equation}
Q\Theta_i \; \longrightarrow \; - \int_0^{\Delta t} dt\, Q \dot\Theta \, .
\end{equation}
The stationary-phase problem is therefore governed not by the original configuration-space Lagrangian $L$, but by its partial Legendre transform with respect to the cyclic coordinate $\Theta$, namely the fixed-$Q$ Routhian
\begin{equation}
R(Y,\dot Y;Q)=L(\Theta,Y;\dot\Theta,\dot Y)-Q\dot\Theta \, ,
\end{equation}
where $\dot\Theta$ is eliminated using Eq.~\eqref{eq:appA_Ptheta}. Solving Eq.~\eqref{eq:appA_Ptheta} then leads to
\begin{equation}
\dot\Theta = \frac{Q-\mathcal A_A[Y]\dot Y^A}{\mathcal I[Y]}\, .
\label{eq:appA_theta_dot_solved}
\end{equation}
Substituting into Eq.~\eqref{eq:appA_Routhian_def}, we obtain
\begin{equation}
R(Y,\dot Y;Q) = \frac12\,G_{AB}[Y]\dot Y^A\dot Y^B - U[Y] - \frac{1}{2\,\mathcal I[Y]} \Big( Q-\mathcal A_A[Y]\dot Y^A \Big)^2\, .
\label{eq:appA_Routhian_explicit}
\end{equation}
This is the exact field-theory counterpart of the reduced fixed-$J$ Routhian in quantum mechanics.

\subsection{Equations of motion at fixed charge}

The reduced action is
\begin{equation}
S_R[Y;Q] = \int_0^{\Delta t} dt\, R(Y,\dot Y;Q)\, .
\label{eq:appA_SR_reduced}
\end{equation}
The fixed-charge equations of motion are therefore the Euler-Lagrange equations for the transverse configuration-space coordinates,
\begin{equation}
\frac{d}{dt}\frac{\delta R}{\delta \dot Y^A} - \frac{\delta R}{\delta Y^A} = 0\, .
\label{eq:appA_Y_eom}
\end{equation}
Using Eq.~\eqref{eq:appA_Routhian_explicit}, this becomes
\begin{align}
0 = \frac{d}{dt} \left[ G_{AB}\dot Y^B + \frac{\mathcal A_A}{\mathcal I} \Big(
Q-\mathcal A_C\dot Y^C \Big) \right] -\frac12\,\partial_A G_{BC}\,\dot Y^B\dot Y^C
+\partial_A U \nonumber\\
\hspace{4em} +\frac{1}{2}\,\partial_A\!\left(\frac{1}{\mathcal I}\right) \Big( Q-\mathcal A_C\dot Y^C \Big)^2 -\frac{1}{\mathcal I} \Big( Q-\mathcal A_C\dot Y^C
\Big)\, \partial_A \mathcal A_B\,\dot Y^B\, ,
\label{eq:appA_Y_eom_explicit}
\end{align}
where $\partial_A\equiv \delta/\delta Y^A$. This is the reduced equation of motion for the fixed-$Q$ tunnelling problem in configuration-space coordinates. The corresponding evolution of the cyclic coordinate is obtained from Eq.~\eqref{eq:appA_theta_dot_solved}.

\subsection{Steadyon and Euclidean formulations}

Up to this point, the discussion in this appendix is independent of the chosen time variable. In the steadyon formulation, the original action is replaced by its regulated version $S_\epsilon$, 
\begin{equation}
S_{R,\epsilon}[Y;Q] = \int dt\, R_\epsilon(Y,\dot Y;Q)\, ,
\label{eq:appA_SR_eps}
\end{equation}
with the regulated Routhian $R_\epsilon$~\eqref{eq:SR_eps_field_local_revised}. 

The steadyon equations are then the Euler--Lagrange equations associated with $R_\epsilon$,
\begin{equation}
\frac{d}{dt}\frac{\delta R_\epsilon}{\delta \dot Y^A} - \frac{\delta R_\epsilon}{\delta Y^A} = 0\, ,
\label{eq:appA_Y_eom_eps}
\end{equation}
together with the fixed-charge relation obtained from the analogue of Eq.~\eqref{eq:appA_Ptheta}.

The leading tunnelling exponent is controlled by the imaginary part of the regulated Routhian action evaluated on the relevant steadyon saddle,
\begin{equation}
\log \Gamma  \sim -2\,\mathrm{Im}\,S_{R,\epsilon}[Y_{\rm st};Q]\, .
\label{eq:appA_decay_steadyon}
\end{equation}
If the initial state carries a non-vanishing energy \(E\), the usual additional term proportional to the Euclidean traversal time is recovered exactly as in the mechanical discussion.

Upon analytic continuation to Euclidean time, the same construction yields the Euclidean Routhian action,
\begin{equation}
S_{E,R}[Y;Q] = \int d\tau\, R_E(Y,\partial_\tau Y;Q)\, .
\label{eq:appA_SER}
\end{equation}
If \(\Theta_E\) denotes the Euclidean continuation of the cyclic coordinate, then
\begin{equation}
S_{E,R} = S_E+\int d\tau\,Q\,\partial_\tau \Theta_E\, .
\label{eq:appA_SER_theta}
\end{equation}
For a configuration with total Euclidean twist
\begin{equation}
\eta = \Delta \Theta_E\, ,
\label{eq:appA_eta_def}
\end{equation}
this becomes
\begin{equation}
S_{E,R}=S_E+\eta Q\, .
\label{eq:appA_SER_etaQ}
\end{equation}
The Euclidean equations of motion are the Euler--Lagrange equations associated with \(R_E\), namely
\begin{equation}
\frac{d}{d\tau}\frac{\delta R_E}{\delta (\partial_\tau Y^A)} - \frac{\delta R_E}{\delta Y^A} = 0\, ,
\label{eq:appA_Y_eom_E}
\end{equation}
and the leading tunnelling exponent takes the form
\begin{equation}
\log \Gamma  \sim -2\,S_{E,R}[Y_I;Q]\, .
\label{eq:appA_decay_E}
\end{equation}

\subsection{Relation to the mechanical case}

The above construction is completely identical to the quantum-mechanical one. The only difference is that, in ordinary mechanics, the reduced coordinates $Y^A$ are finite in number, while in field theory they label an infinite-dimensional configuration space. In this sense, quantum mechanics is simply the $0+1$-dimensional version of the same structure.

For instance, in $0+1$ dimensions, a single complex scalar reduces to an ordinary particle in two real dimensions,
\begin{equation}
\phi(t)=\frac{r(t)}{\sqrt2}e^{i\varphi(t)}\, ,
\label{eq:appA_01d_scalar}
\end{equation}
with
\begin{equation}
L=\frac12\dot r^2+\frac{r^2}{2}\dot\varphi^2-V(r^2/2)\, , \qquad J=\frac{\partial L}{\partial \dot\varphi}=r^2\dot\varphi\, .
\label{eq:appA_01d_mech}
\end{equation}
The reduced fixed-$J$ Routhian is then obtained exactly as above by eliminating $\dot\varphi$ in favour of $J$. The field-theory construction presented here is therefore not merely analogous to the mechanical one, but its exact infinite-dimensional generalisation.

\bibliography{ref}

\end{document}